\documentclass[9pt,twocolumn,twoside]{pnas-new}
\usepackage[version=3]{mhchem}
\usepackage{fixltx2e}
\templatetype{pnasresearcharticle} 
\usepackage{graphicx}
\usepackage{caption}
\usepackage{balance}
\usepackage{subcaption}
\usepackage{cleveref}

\title{Tomonaga-Luttinger liquid and quantum criticality in spin-$\frac{1}{2}$ antiferromagnetic Heisenberg chain \ce{C14 H18 Cu N4 O10} via Wilson ratio}

\author[a]{Sharath Kumar Channarayappa}
\author[b]{Sankalp Kumar}
\author[c]{N. S. Vidhyadhiraja}
\author[b]{Sumiran Pujari}
\author[d]{M. P. Saravanan}
\author[a]{Amal Sebastian}
\author[e]{Eun Sang Choi}
\author[e]{Shalinee Chikara}
\author[a]{Dolly Nambi}
\author[a]{Athira Suresh}
\author[f]{Siddhartha Lal}
\author[a,$\dagger$]{D. Jaiswal-Nagar}

\affil[a]{School of Physics, IISER Thiruvananthapuram, Vithura, Kerala-695551, India}
\affil[b]{Department of Physics, Indian Institute of Technology Bombay, Mumbai-400076, India}
\affil[c]{Theoretical Sciences Unit, Jawaharlal Nehru Center for Advanced Scientific Research, Jakkur, Bengaluru-560064, India}
\affil[d]{UGC-DAE Consortium for Scientific Research, University Campus, Khandwa Road, Indore-452001, India}
\affil[e]{National High Magnetic Field Lab (NHMFL), Tallahassee, Florida 32310, USA}
\affil[f]{Department of Physical Sciences, Indian Institute of Science Education and Research-Kolkata, Mohanpur Campus, West Bengal-741246, India}

\leadauthor{Sharath}

\significancestatement{
This work presents an experimental phase diagram of a spin-$\frac{1}{2}$ antiferromagnetic Heisenberg chain, utilizing Wilson ratio as a robust indicator of quantum criticality in Tomonaga-Luttinger liquids. Through comprehensive measurements in single crystals of novel material \ce{C14H18CuN4O10}, distinct phases of Tomonaga-Luttinger liquid, quantum critical and fully polarized phase are elucidated across a wide parameter space. Very good agreement between theory and experiment validates \ce{C14H18CuN4O10} as an exceptional realization of spin-$\frac{1}{2}$ antiferromagnetic Heisenberg chain system. Our findings highlight the broader impact of systems with low saturation fields in studying quantum critical phenomena and underscore the significance of Wilson ratio in characterizing such systems. This work advances understanding in one-dimensional quantum systems and opens avenues for exploring emergent quantum phenomena across diverse materials.}

\authorcontributions{Author contributions: D.J.-N. planned the research; S.K.C., S.K., N.S.V., S.P., M.P.S., A.S., E.S.C., S.C., A.S., D.N., S.L. and D.J.-N. performed research; S.K.C. and D.J.-N. analyzed data and wrote the manuscript with inputs from all co-authors.}

\authordeclaration{There are no conflicts of interest to declare.}
\correspondingauthor{\textsuperscript{$\dag$}To whom correspondence should be addressed. E-mail: deepshikha@iisertvm.ac.in}

\keywords{Quantum criticality $|$ Tomonaga-Luttinger liquid $|$ Wilson ratio $|$ One-dimensional antiferromagnetic Heisenberg chain $|$ Quantum phase transition} 

\begin{abstract}
	The ground state of a one-dimensional spin-$\frac{1}{2}$  uniform antiferromagnetic Heisenberg chain (AfHc) is a Tomonaga-Luttinger liquid which is quantum-critical with respect to applied magnetic fields upto a saturation field $\mu_0 H_s$ beyond which it transforms to a fully polarised state. Wilson ratio has been predicted to be a good indicator for demarcating these phases [Phys. Rev. B 96, 220401 (2017)]. From detailed temperature and magnetic field dependent magnetisation, magnetic susceptibility and specific heat measurements in a metalorganic complex and comparisons with field theory and quantum transfer matrix method calculations, the complex was found to be a very good realisation of a spin-$\frac{1}{2}$ AfHc. Wilson ratio obtained from experimentally obtained magnetic susceptibility and magnetic contribution of specific heat values was used to map the magnetic phase diagram of the uniform spin-$\frac{1}{2}$ AfHc over large regions of phase space demarcating Tomonaga-Luttinger liquid, saturation field quantum critical, and fully polarised states. Luttinger parameter and spinon velocity were found to match very well with the values predicted from conformal field theory.
\end{abstract}

\begin{document}
	
	\maketitle
	\thispagestyle{firststyle}
	\ifthenelse{\boolean{shortarticle}}{\ifthenelse{\boolean{singlecolumn}}{\abscontentformatted}{\abscontent}}{}

\begin{figure*}[hbt!] 
	\centering
	\includegraphics[width=1\linewidth]{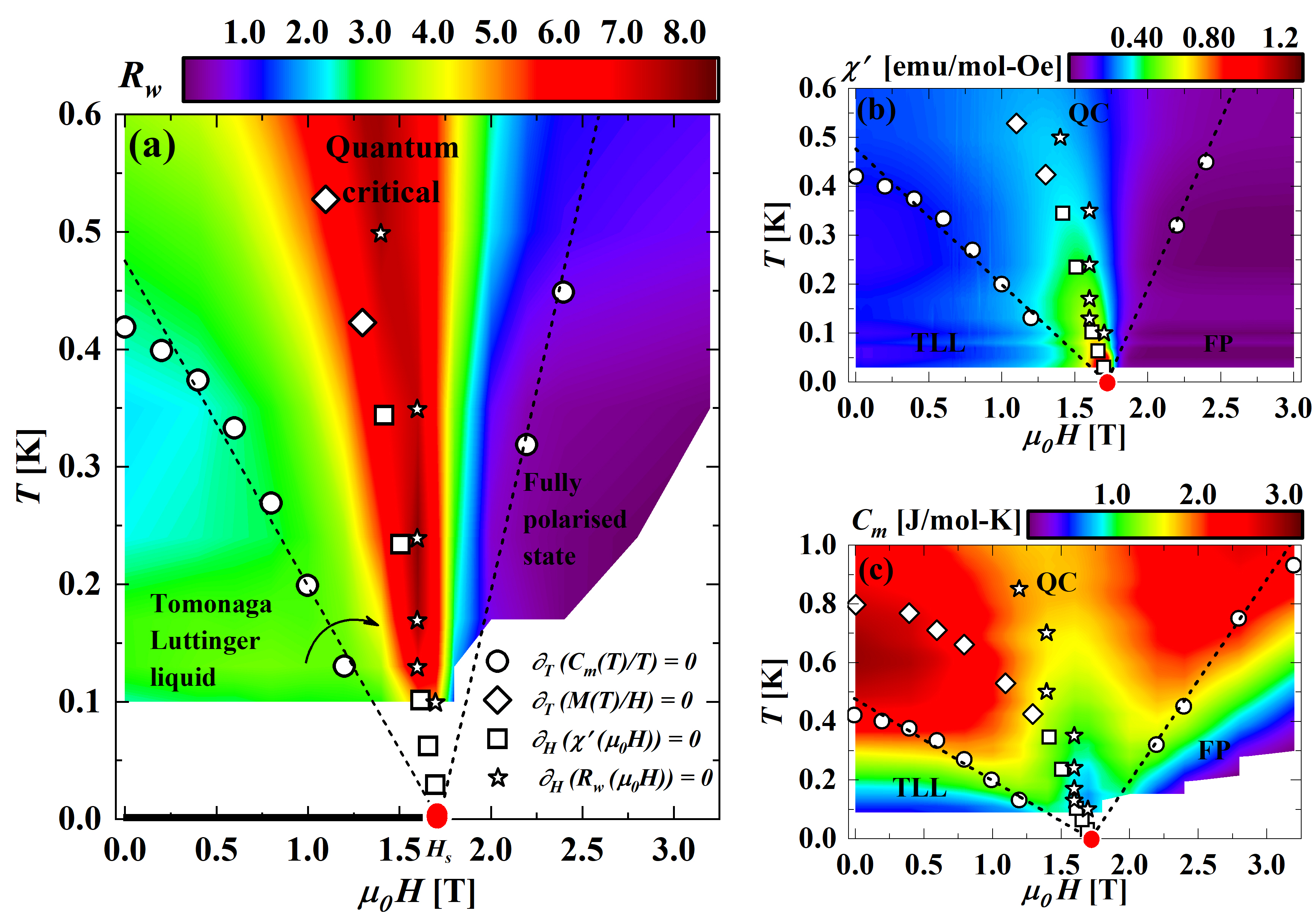}
	\caption{(a) Temperature-magnetic field phase diagram of CuD displaying Tomonaga-Luttinger liquid, quantum critical cone and Fully polarised phases. The $T = 0$ K line of critical points has been shown by a thick black line that terminates in a quantum critical saturation field end point, $\mu_0 H_s$, shown by a red filled dot. Wilson ratio, $R_w(\mu_0 H,T)$, is shown on the contour plot with the colour scale shown at the top where dark red denotes the highest value and dark blue denotes the least value. Open circles represent the maxima of $C_m(T)/T$ while the maxima of the $M(T)/H$ and $\chi'(\mu_0 H)$ are shown by open diamonds and open squares respectively. The peak in $R_w(\mu_0 H)$ (see SI Fig. S5) is shown as open stars. The black  dashed lines represent fits to equation \ref{eqn:QC line} with $\alpha_1$ and $\alpha_2$ as fit parameters which were obtained as 3.61 and 1.45 respectively. Curved arrow represents the cross-over from $z = 1$ to $z = 2$ state. (b) and (c) represent the phase diagram made using $\chi'(\mu_0 H,T)$ and $C_m(\mu_0 H,T)$ respectively. In (b), 15 scans of temperature with field dependent $\chi'(\mu_0 H)$ ($\mu_0 H \perp$ a-axis)  were used to create the contour plot. In (c), 16 scans of magnetic fields with temperature dependent $C_m(T)$  ($\mu_0 H \perp$ a-axis) were used to create the contour plot.} 
	\label{fig:phase diagram}
\end{figure*}

\begin{figure*}
	\centering
	\includegraphics[width=1\linewidth]{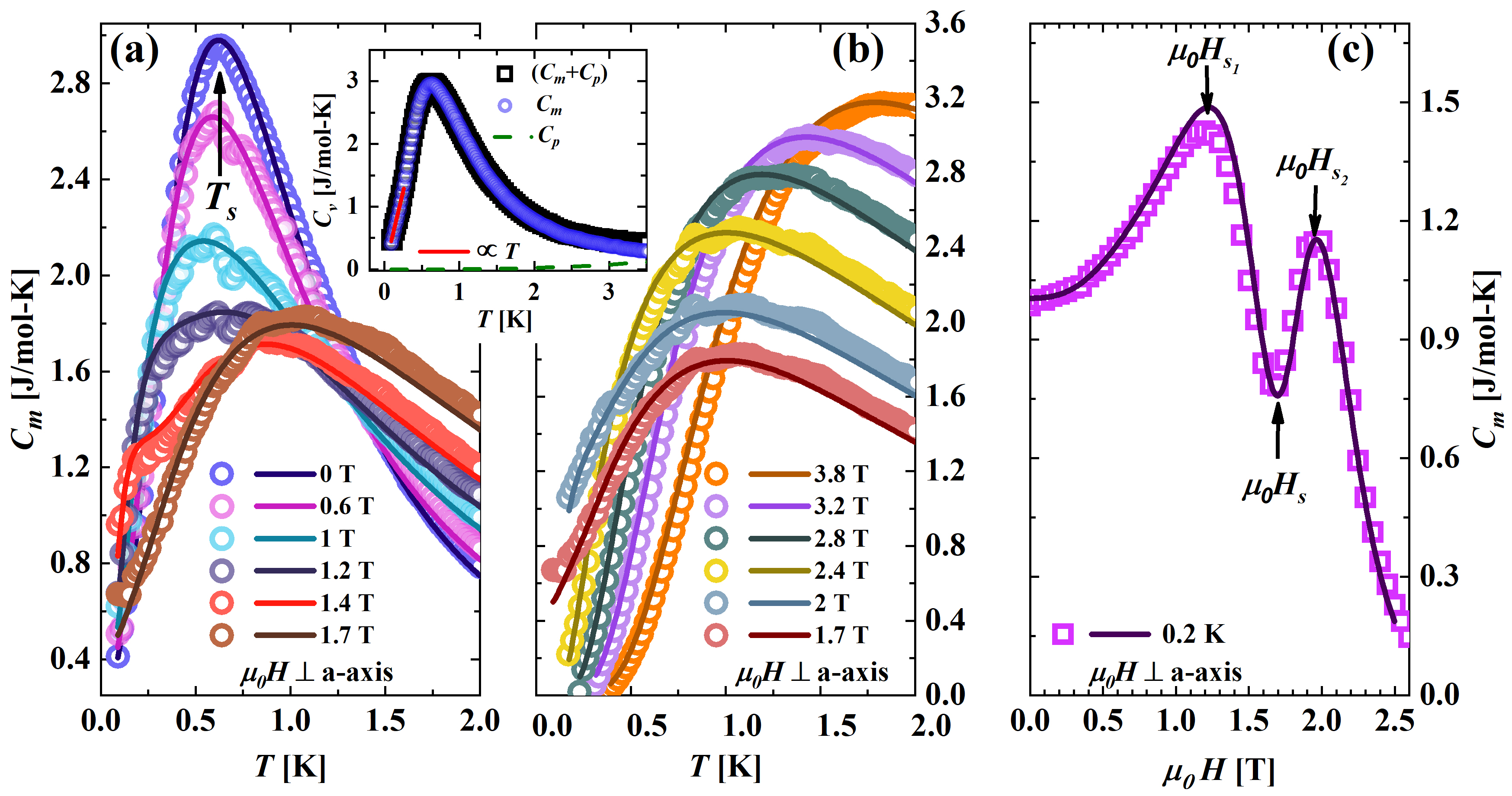}%
	\caption{Black open squares in inset of (a and b) represent the actual measured $C_v(T)$ while the green dashed line represents the estimated $C_p(T)$ to low temperatures from the high temperature Debye-Einstein model fit. $C_m(T)$, illustrated as blue open circles, is obtained by subtracting  $C_p(T)$ from $C_v(T)$. At low temperatures, red solid line is a linear fit to $C_m(T)$ obtained for $\mu_0 H = 0$ T. (a) Open coloured circles represent the temperature variation of $C_m(T)$ at various applied magnetic fields up to $\mu_0 H_s$, and (b) at $\mu_0 H_s$ and above. The corresponding QTM calculations are represented by solid curves of the same colour. (c) Purple open circles represent the magnetic field variation of $C_m$ at $T = 0.2 $ K while the solid line is the result of QTM calculations. Arrows indicate the position of the two maxima and a QCP.}
	\label{fig:specific heat}
\end{figure*}

\dropcap{P}	Phase diagrams provide a comprehensive means to study the complex behavior of systems near phase transitions, bridging theoretical predictions with experimental observations and enhancing our understanding of emergent phenomena. A text book example is the pressure-temperature phase diagram of water that has a line of first order phase transitions that terminate into a critical point of $P_c$ = 221 bar and $T_c$ = 374$^{\circ}$C \cite{chaikin,wagner}. An analogous phase diagram was recently observed in a frustrated two-dimensional spin-$\frac{1}{2}$ magnet, \ce{SrCu2(BO3)2} \cite{jimenez}. When the finite temperature critical point is suppressed to T = 0 K, a quantum critical point (QCP) associated with a quantum phase transition (QPT) emerges \cite{vojta,sachdev}. A QCP is expected to affect a finite portion of the phase diagram in a cone-like region bounded by the entanglement temperature \cite{mathew}. However, very few examples of phase diagrams exhibiting the quantum critical cone along-with the phases associated with the QCP, exist in the literature, the prominent being the phase diagram of the heavy fermion \ce{YbRh2Si2} \cite{gegenwart}. The cone-like quantum critical (QC) region is demarcated by boundaries that are determined by the condition $k_BT \propto |{r-r_c}|^{\nu z}$, where $\nu$ and $z$ denote the correlation length critical exponent and dynamic critical exponent respectively that are usually universal \cite{vojta,sachdev}. The non-thermal control parameter $r$ that is used to tune the QCP are pressure, doping, magnetic field etc. In this regard, magnetic field turns out to be a very useful handle to probe quantum criticality in diverse systems due to the ease of application of a magnetic field to reversibly and continuously tune a system towards a QCP \cite{gegenwart, sebastian, tanatar, ruff, das}. However, large values of magnetic fields needed to tune a QCP in systems having large exchange coupling constants make experimental investigations of phase diagrams difficult in quantum critical systems \cite{lake,motoyama,kono,breunig}. Therefore, systems having low values of exchange coupling constants in which the QCP could be tuned by low values of applied magnetic fields easily accessible by laboratory magnets provide excellent platforms using which the complexities associated with QPT's could be studied and phase diagrams made. Our work describes a detailed and extended phase diagram of a quantum critical system with quite favourable coupling constant.\\ 
QPT's have been observed in diverse and complex systems ranging from heavy fermions to high temperature superconductors \cite{gegenwart,subir,kuo} but have not yet been understood completely. A study of QPT's in systems that are realisations of exactly solvable models can offer deeper insights into this phenomenon. One such exactly solvable model is the spin-$\frac{1}{2}$ one-dimension (1D) antiferromagnetic Heisenberg chain (AfHc) that can be described by a relativistic field theory in the low energy limit. Specifically, the Tomonaga-Luttinger liquid (TLL) theory, a relativistic free boson field theory, is known to describe the ground state of a spin-$\frac{1}{2}$ AfHc \cite{klumper}. In this theory, the velocity of the spin-waves $u$ is a variable unlike the speed of light in a true relativistic theory which is a constant \cite{maeda}. This and another parameter of the TLL theory, namely, the Luttinger parameter $K$, can be determined in an integrable model \cite{giamarchi} and can fully describe the low-energy features within the TLL framework. For free fermions $K=1$, while $K<1$ and $K>1$ represent repulsive and attractive interactions, respectively. It is anticipated that $K$ would change continuously for the simple spin-$\frac{1}{2}$ AfHc, from $K=0.5$ in the zero field to $K=1$ at saturation magnetic field $\mu_0 H_s$ \cite{giamarchi}.\\
The Hamiltonian of the spin-$\frac{1}{2}$ AfHc in a magnetic field $H$ is given by:
\begin{equation}
	\mathcal{H} = J\sum_{i} \vec{S}_i \cdot \vec{S}_{i+1} - g\mu_{B}H\sum_{i} \vec{S}_{i}^{z}
	\label{eq:Hamiltonian}
\end{equation}
where $J$ is the exchange coupling constant in the chain. At $T=$ 0 K and $\mu_0H=$ 0 T, the spin-$\frac{1}{2}$ AfHc fails to develop any long range order, however, the correlation functions exhibit an algebraic decay making the AfHc a quantum critical system. The spin-$\frac{1}{2}$ AfHc is also critical with respect to an applied field up-to the saturation field $H_s$, given by $g\mu_Bm_sH_s=J$ \cite{muller} above which it transforms to a fully polarised (FP) state as shown in Fig. \hyperref[fig:phase diagram]{1} which is an exact eigenstate of the Hamiltonian equation \ref{eq:Hamiltonian} different from TLL. So, $H_s$ marks the end-point of a line of QCP's separating the fully polarised state from the partially magnetised TLL state at lower fields \cite{wolf}. The excitation's of the TLL are spinons that are topological excitation's in the spin order and form a continuum excitation spectrum over an extended range of energy and momentum \cite{lake}. In contrast, the excitation spectrum of the FP state comprises gapped excitation's \cite{he}.\\
Wilson ratio, $R_w$, defined as the ratio of magnetic susceptibility $\chi'$ to specific heat $C$ divided by temperature $T$ \cite{sommerfeld,wilson}:
\begin{equation}
	R_w = \frac{4}{3}\bigg(\frac{\pi k_B}{g\mu_{B}}\bigg)^2\frac{\chi'}{C / T}
	\label{eqn:Wilson ratio}
\end{equation}
is proposed to be an important parameter for characterising the TLL phase boundary as well as the QC region associated with the saturation field critical point \cite{he}. In equation \ref{eqn:Wilson ratio}, $k_B$, $\mu_{B}$ and $g$ denote Boltzmann constant, Bohr magneton and Land\'{e} factor respectively. $R_w = 1$ for a system of non-interacting fermions \cite{hewson} and equals 2 in the Kondo regime of the impurity problem \cite{wilson}. Wilson ratio quantifies interaction effects and spin fluctuations in a strongly correlated system \cite{guan,johnston}. For instance, $R_w$ was found to have a value close to 8 in the vicinity of the QCP in a strongly correlated layered cobalt oxide \ce{BiBa_{0.66}K_{0.36}O2CoO2} \cite{limellete}. Even though the magnetic field variation of $R_w$ was studied in this work, an experimental phase diagram using $R_w$ as contours was not made. Wilson ratio is expected to be in the range $0 < R_w < 10$ across the phase diagram associated with a TLL \cite{he}. \\
There are several materials, for instance \ce{KCuF3} and \ce{Sr2CuO3}, that are good realisations of a spin-$\frac{1}{2}$ AfHc. However, their large values of exchange coupling constants, $J/k_B \sim$ 380 K in \ce{KCuF3} \cite{lake,hirakawa,nagler,tennant1993,tennant1995} and 2200 K in \ce{Sr2CuO3} \cite{motoyama}, make the corresponding saturation field $\mu_0H_s$ extremely high at hundreds and thousands of Teslas. The most well-studied spin-$\frac{1}{2}$ AfHc is \ce{Cu.(C4H4N2).(NO3)2} (CuPzN) that has a moderate $J/k_B$ of $\sim$ 10.3 K \cite{hammar,kono,breunig} making $\mu_0 H_s \sim$ 14 T. This makes the investigation of the fully polarised state in CuPzN difficult using laboratory magnets. Additionally, CuPzN undergoes a 3D ordering at 0.107 K \cite{lancaster} complicating the investigation of the thermodynamics of the spin-$\frac{1}{2}$ AfHc.\\
In this work, we have investigated the thermodynamics of the spin-$\frac{1}{2}$ AfHc over a large range of parameters from $\mu_0 H$ = 0 T till saturation field $\mu_0 H_s$ and far above $\mu_0 H_s$ (FP state) in a new spin-$\frac{1}{2}$ AfHc compound copper bisoxalate amminopyridate \ce{(C5 H7 N2)2 [Cu(C2 O4)2] .2 (H2 O)} having the formula \ce{C14H18CuN4O10} and abbreviated as CuD, that has a $J/k_B$ value of $\sim$ 1.23 K and a corresponding saturation field $\mu_0 H_s$ of $\sim$ 1.7 T \cite{kumar}. Magnetisation normalised by field $M(T)/H$, magnetic susceptibility $\chi'(\mu_0H)$ and magnetic contribution of specific heat $C_m(T)$ curves generated using quantum transfer matrix calculations are found to fit the experimentally produced data excellently over a large range of temperatures and field values. Magnetic susceptibility and magnetic contribution of specific heat were used to calculate the Wilson ratio using which a complete phase diagram of a spin-$\frac{1}{2}$ Heisenberg antiferromagnetic chain has been drawn experimentally for the first time marking the Tomonaga-Luttinger liquid, quantum critical and fully polarised phases. The quantum critical phase boundaries corresponding to the saturation field quantum critical point were found to extend to a large portion of the phase diagram unlike CuPzN where deviations were observed due to a 3D ordering. We observed excellent data collapse in thermodynamic properties arising due to the quantum critical scaling. Finally, we obtained spinon velocity $u$ and the Luttinger parameter $K$ experimentally whose values were found to match perfectly with theoretical predictions \cite{giamarchi}. The sign of the parameter $K$ revealed that the strength and nature of spinon interactions are repulsive, whereas the spinon velocity $u$ revealed that the dynamics are low-energy. With a $J/k_B$ of 0.106 meV, the spinon bandwidth is expected to be $\sim$ 0.16 meV in CuD. This small value of the bandwidth and observations from the transport measurements which indicate that CuD is a robust insulator, the charge channel with holon excitation is expected to be negligible in CuD.\\
CuD crystallises in the monoclinic crystal structure (P21/c) with lattice parameters $a=3.7064$ $\AA$, $b=20.2976$ $\AA$ and $c=11.9059$ $\AA$. The structure comprises \ce{Cu^{2+}} ions that coordinate with four \ce{O} atoms in the basal plane to form a square planar geometry. The \ce{[Cu(C2O4)]^{-2}} ions link together, forming a straight \ce{Cu^{2+}} chain along the crystallographic $a$-axis as shown in SI Fig. S1. This chain has a regular \ce{Cu}$\cdots$\ce{Cu} spacing of 3.706 $\AA$, formed by corner-sharing oxygen atoms of the \ce{CuO6} octahedra \cite{kumar}. To investigate the thermodynamics of CuD, external magnetic field was applied perpendicular to the chain direction. As is well-known, the excitation spectrum of a TLL comprises multispinons continuum that is isotropic in the absence of any field \cite{giamarchi,affleck,klumper,klumper2000}. However, on field application, the spin correlations become anisotropic, such that S$_{xx}(\textbf{q},w)$ = S$_{yy}(\textbf{q},w) \neq$ S$_{zz}(\textbf{q},w)$. Therefore, the maximum effect to the total spectrum arises from the tranverse contributions where a significant loss of spectral weight arises at the centre of the Brillouin zone when compared to the longitudinal spectrum \cite{halg}. Accordingly, the field was applied perpendicular to the chain direction ($ \mu_0H \perp$ crystallographic $a$-axis) since the thermodynamics of a spin-$\frac{1}{2}$ AfHc in a field is determined mainly by the transverse excitation modes \cite{xiang,hagiwara, halg}.\\
The thermodynamics of the integrable spin-$\frac{1}{2}$ AfHc was calculated by considering a quasi-linear energy-momentum dispersion of the spinons having a velocity $u$ using which the free energy equations are solved analytically and implemented numerically using the quantum transfer matrix (QTM) method (see SI section \hyperref[Supplementary]{6} for details). The field-theoretic spinless fermionic description of the TLL gives the following scaling form of the free energy
\begin{equation}
	F(T,\mu) = \frac{J - 2 g \mu_B H}{4} + \frac{(k_B T)^{3/2}}{\sqrt{J}} \Phi \left( \frac{g \mu_B (H_s - H)}{k_B T}\right)
	\label{eqn:free energy}
\end{equation} 
at leading order. The first term is the ground state energy denoting the fully polarised state while the second term denotes the asymptotic scaling behaviour close to the QCP arising from the thermal population of spinons. Higher order corrections come from spinon interactions that are irrelevant in the renormalisation group sense and will be sub-dominant.
The magnetic phase diagrams of CuD produced using Wilson ratio ($R_w$), magnetic susceptibility ($\chi'$) and magnetic contribution of the specific heat ($C_m$) are shown in Figs. \hyperref[fig:phase diagram]{1a-1c} respectively. The contour plot made using $R_w$, distinguishes three distinct regions: a gapless Tomonaga-Luttinger liquid region, a quantum critical cone and the gapped fully polarised region. The ground state wave-function of the TLL containing entangled spins evolves continuously with the application of field such that at the saturation field $\mu_0 H_s$, the TLL breaks down and an unentangled fully polarised state appears on the right side of $\mu_0 H_s$ \cite{mathew}. It has been reported that Wilson ratio and specific heat is a better marker of the TLL phase boundary \cite{he} compared to the magnetisation/magnetic susceptibility measurements. Accordingly, in Fig. \hyperref[fig:phase diagram]{1a} the region below the open circles between $0 \le \mu_0 H \le \mu_0 H_s$ marked by maxima in $C_m(T)/T$ (see SI Fig. \hyperref[Supplementary]{S4a}) has been marked as the TLL phase. The contour plots of the Wilson ratio suggest that the TLL phase is subdivided into two regions (see Fig. \hyperref[fig:phase diagram]{1a}), which is not evident from the susceptibility contours of the same color, Fig. \hyperref[fig:phase diagram]{1b}, but can be inferred from the specific heat contours that also exhibit gradients in values within the TLL phase (Fig. \hyperref[fig:phase diagram]{1c}).\\
The fully polarised state is also marked by specific heat measurements as temperatures in $C_m(T)/T$ plots (see SI Fig. \hyperref[Supplementary]{S4b}) where an exponential drop occurs in the specific heat. From Fig. \hyperref[fig:phase diagram]{1a}, it can be seen that the Wilson ratio $R_w$ falls to values less than 0.5 in this region. Magnetisation also falls to a low value in this region (see Fig. \hyperref[fig:phase diagram]{1b}), while specific heat remains finite and with variable magnitude (see Fig. \hyperref[fig:phase diagram]{1c}). However, the Wilson ratio has a small and uniform contour in the FP state, demonstrating the usefulness of the Wilson ratio as an indicator for mapping the phase diagram of a quasi-one dimensional spin-$\frac{1}{2}$ Heisenberg antiferromagnet.\\
The universal scaling behaviour associated with the quantum critical endpoint $\mu_0 H_s$ and signaling a breakdown of the TLL phase \cite{he2020} is captured by the quantum critical phase boundaries. These phase boundaries are given by the expression \cite{he}:
\begin{equation}
	T_{spinon} = \frac{g \mu_{B} \mu_0 }{\alpha_1k_B}(H_s - H);~~~~~ T_{magnon} = - \frac{g \mu_{B} \mu_0}{\alpha_2k_B}(H_s - H)   
	\label{eqn:QC line}
\end{equation}
where $T_{spinon}$ and $T_{magnon}$ represent the left and right lines of the quantum critical cone governed by the spinon excitations of the TLL phase and magnon excitations of the fully polarised phase respectively. $\alpha_1$ and $\alpha_2$ are constants. A comparison of the equation \ref{eqn:QC line} with the quantum critical phase boundaries $k_BT \propto |{r-r_c}|^{\nu z}$ yields $\nu z = 1$. It is known that correlation-length exponent $\nu = \frac{1}{2}$ and dynamical exponent $z = 2$ at the saturation field $\mu_0 H_s$ of a spin-$\frac{1}{2}$ AfHc \cite{klumper,he,klumper2000}. So, the phase boundaries associated with the saturation field critical point of a spin-$\frac{1}{2}$ AfHc is governed by the dynamic critical exponent $z = 2$. The theoretical phase boundaries (black dashed lines of Fig. \hyperref[fig:phase diagram]{1a}) are found to follow the experimental data points in CuD to a very large portion of the phase diagram and not limited to areas close to $\mu_0 H_s$, in contrast to CuPzN \cite{jeong2013,he} where deviations are observed due to 3D ordering in CuPzN. Additionally, since in the TLL phase $z = 1$ \cite{sachdev}, the left QC line governs the transformation of $z = 1$ to $z = 2$ as shown by thick curved arrow in Fig. \hyperref[fig:phase diagram]{1a}. On the other hand, the right QC line demonstrates the transformation of the field induced gap, $\Delta$, that goes linearly with field, $\Delta \propto \mu_0 (H-H_s)$ \cite{muller}, to the quantum critical region.
\begin{figure*}
	\centering
	\includegraphics[width=1\linewidth]{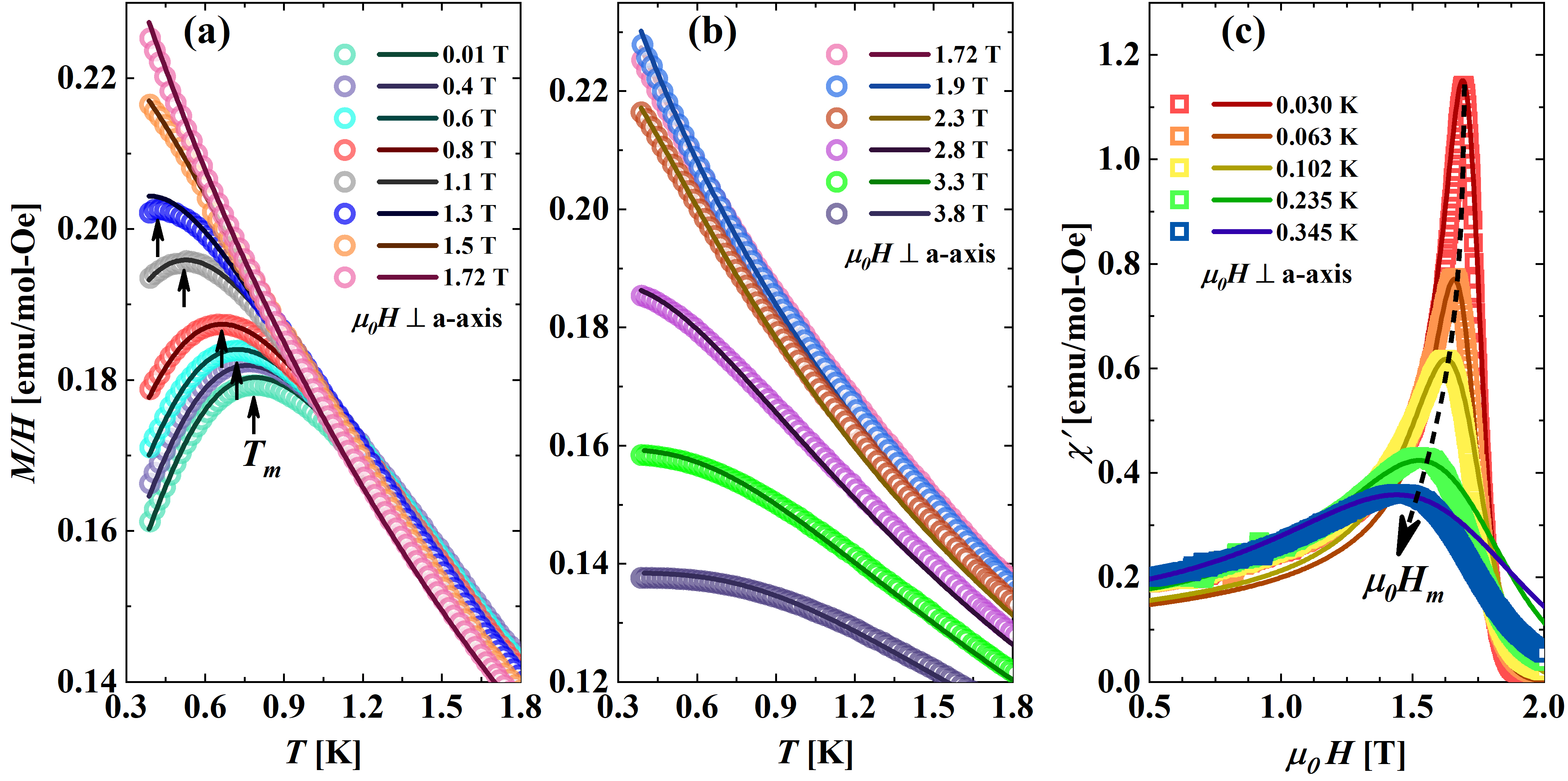}%
	
	\caption{(a) At various fields upto $\mu_0 H_s$, $M(T)/H$, is shown by open coloured circles. The corresponding coloured solid lines show the quantum transfer matrix theoretical calculations. The peak temperature denoted by $T_m$ is represented by black arrows. (b) Coloured open circles and corresponding coloured solid lines represent $M(T)/H$ and quantum transfer matrix calculations respectively at saturation field $\mu_0 H_s$ and above. (c) Open triangles represent the field variation of  $\chi'(\mu_0 H)$, at different values of temperatures while the solid lines of the same colour denote the result of QTM calculations. Black dashed line is a guide to the eye and indicates the shift of the maxima.}
	\label{fig:magnetisation}
\end{figure*}
Magnetisation, magnetic susceptibility and specific heat measurements used to make the phase diagram of Fig. \hyperref[fig:phase diagram]{1} are shown in Figs. \hyperref[fig:magnetisation]{2} and \hyperref[fig:specific heat]{3}. Coloured open circles in Figs. \hyperref[fig:magnetisation]{2a} and \hyperref[fig:magnetisation]{2b} depict the magnetic field evolution of temperature dependent magnetisation normalised to field, $M(T)/H$ at fields below (Fig. \hyperref[fig:magnetisation]{2a}) and above (Fig. \hyperref[fig:magnetisation]{2b}) the saturation field $\mu_0 H_s$. The corresponding solid lines in Figs. \hyperref[fig:magnetisation]{2a} and \hyperref[fig:magnetisation]{2b} denote the result of QTM calculations done with only one free parameter, $J$. The calculated $M(T)/H$ curves were obtained as a function of $T/J/k_B$. It can be seen that the calculated curves fall exactly on top of the experimentally obtained ones once the temperature axis is scaled by the experimentally obtained value of $J/k_B = 1.23 $ K indicating that CuD is an excellent realisation of a spin-$\frac{1}{2}$ AfHc. The low field curves are characterised by a peak demarcated as $T_m$ (open diamonds of Fig. \hyperref[fig:phase diagram]{1}) indicating the cross-over from a TLL to QC. $T_m$ is predicted to occur at $T_m\sim 0.641J$ with the maximum value $(M(T)/H)_{max}\sim 0.146 Ng^2\mu_{B}^{2}/J$ in the limit of $\mu_0 H\rightarrow0$ \cite{bonner,johnston}. From the experimentally obtained value of $T_m = 0.79$ K and $(M(T)/H)_{max} = 0.18$ emu/mol*Oe at the lowest applied field of 0.01 T, $J/k_B$ is obtained as 1.23 K while the Land\'{e} g-factor, $g$, is obtained as 2.03. These numbers exactly match the values obtained from fitting the uniform spin-$\frac{1}{2}$ AfHc model (high temperature series expansion) to the 0.01 T data (see SI section \hyperref[Supplementary]{2} and Fig. \hyperref[Supplementary]{S2}). Using $J/k_B = 1.23$ K and $g = 2.03$, $\mu_0 H_s$ is obtained as 1.8 T. From Fig. \hyperref[fig:magnetisation]{2a}, it is to be noted that as the applied field increases, $T_m$ steadily decreases such that for a field of 1.5 T, it falls below the lowest measurable temperature of 0.38 K. At fields near $\mu_0 H_s$, $M(T)/H$ exhibits diverges as $T\rightarrow0 $ K, indicating quantum criticality \cite{klumper,xiang}. For fields above $\mu_0 H_s$, the magnetisation plateaus at low temperatures where the ground state is a gapped, field-induced polarised state \cite{klumper,xiang}.\\
For fields at which $T_m$ fell below the measurable temperatures of 0.38 K (lowest attainable temperature of SQUID magnetometer), it was difficult to ascertain the TLL phase boundary, especially for fields closer to $\mu_0 H_s$. To overcome this problem, field dependent magnetic susceptibility $\chi'(\mu_0 H)$ measurements were performed at different temperatures upto 30 mK using an ac susceptometer  and shown as coloured open squares in Fig. \hyperref[fig:magnetisation]{2c}. The curves present a peak at magnetic field $\mu_0 H_{m}$ (open squares in Fig. \hyperref[fig:phase diagram]{1}) while the corresponding solid curves are the result of QTM calculations obtained in the way described above. As can be observed, the match between the calculated theory curves and the experimentally obtained curves is very good. The peak magnetic field $\mu_0H_{m}$ is found to shift to lower magnetic fields with an increase in temperature due to increased thermal fluctuations.
The contour plot of Fig. \hyperref[fig:phase diagram]{1c} was made using specific heat measurements shown in Figs. \hyperref[fig:specific heat]{3a} and \hyperref[fig:specific heat]{3b}. Temperature dependence of magnetic specific heat, $C_m(T)$, shown in Figs. \hyperref[fig:specific heat]{3a} and \hyperref[fig:specific heat]{3b} was obtained by subtracting the phonon contribution, $C_p(T)$, shown as green dashed line in the inset of Figs. \hyperref[fig:specific heat]{3a} and \hyperref[fig:specific heat]{3b} from the total specific heat $C_v(T)$ (black open squares in inset of Fig. \hyperref[fig:specific heat]{3a} and \hyperref[fig:specific heat]{3b}). $C_p(T)$ was calculated using the Debye-Einstein model having one and two terms respectively (see SI Section \hyperref[Supplementary]{3} and Fig. \hyperref[Supplementary]{S3} for details). Although phonons start to contribute at 1.2 K and above (green dashed line), the nuclear contribution to $C_m(T)$ is relevant only at ultra-low temperatures of 100 mK and below as well as high magnetic fields of the order of 10 T, and hence, can be neglected in our measured range of temperatures and magnetic fields. The blue open circles in the insets of Figs. \hyperref[fig:specific heat]{3a} and \hyperref[fig:specific heat]{3b} represent the zero field $C_m(T)$, which indicates the absence of long-range ordering (LRO) down to 100 mK. Temperature variation of $C_m(T)$ is shown in Figs. \hyperref[fig:specific heat]{3a}-\hyperref[fig:specific heat]{3b} for various values of applied magnetic field upto saturation field $\mu_0 H_s$ (Fig. \hyperref[fig:specific heat]{3a}) and $\mu_0 H_s$ and beyond (Fig. \hyperref[fig:specific heat]{3b}). Dark coloured solid lines of the same colour as the open circles show the result of QTM calculations obtained using a single free parameter, $J$. $C_m(T)$ was obtained as a function of $T/J/k_B$ and scaled to the temperature axis by normalising it with the experimentally obtained $J/k_B$. As with the magnetisation results above, the match between theoretically obtained and experimentally obtained curves is excellent reiterating that CuD is an excellent realisation of a spin-$\frac{1}{2}$ AfHc. The curves reveal a maximum in $C_m(T)/T$ as the temperature is varied indicating a transformation from a TLL phase to a QC phase (see Fig. \hyperref[fig:phase diagram]{1}). At 0 T, the peak temperature of $M(T)/H$ is nearly twice that of $C_m(T)/T$ since $M(T)/H$ measures two-particle excitations while $C_m(T)/T$ measures one-particle density of states \cite{xiang}. In Fig. \hyperref[fig:specific heat]{3a} maximum in $C_m(T)$ at $ \mu_0H = 0 $ T is predicted to arise at $T_{s} = 0.48J/k_B$ \cite{johnston}. Using the value of $J/k_B = 1.23$ K obtained from magnetisation data above, $T_{s}$ should be at 0.59 K. It is satisfying to note that $T_{s} = 0.6 $ K, in very good agreement with theoretical predictions.\\
Figs. \hyperref[fig:specific heat]{3a} and \hyperref[fig:specific heat]{3b} present striking differences in the low temperature behavior of $C_m(T)$ when the applied external magnetic field is below or above the saturation field $\mu_0 H_s$ respectively. While for $\mu_0 H < \mu_0 H_s$, $C_m(T)$ at low temperatures is finite indicating a finite density of low energy spinons, for $\mu_0 H > \mu_0 H_s$, the low temperature $C_m(T)$ falls to zero exponentially due to the opening up of a gap in the density of states after the transformation of the TLL state to that of a FP state. Furthermore, $C_m(T)$ increases linearly with temperature indicating the fermionic nature of the quasiparticles (which are spinons) of the TLL state. With an increase in the applied field, the maximum value of specific heat, $(C_m(T))_{max}$ reduces in magnitude. The application of field results in the generation of spin strings of different lengths in the highly entangled ground state reducing the ground state's number of spinons at the Fermi points \cite{he,bera}, and consequently, the specific heat. Fig. \hyperref[fig:specific heat]{3c} depicts the field variation of $C_m(\mu_0H)$ at the lowest measured temperature of 0.2 K. From the plot, one can observe the presence of two maxima at $\mu_0H_{s_1}$ and $\mu_0H_{s_2}$ indicating the transformation from the TLL to a QC (at $\mu_0H_{s_1}$) and that to a fully FP state (at $\mu_0H_{s_2}$) respectively. So, at $\mu_0H_{s_1}$, quantum and thermal fluctuations reach an equal footing while at $\mu_0H_{s_2}$, fluctuations due to magnon dominate. The intermediate region between $\mu_0H_{s_1}$ and $\mu_0H_{s_2}$ denote saturation field quantum critical region where thermal fluctuations strongly couple with quantum fluctuations \cite{he}. The solid line in Fig. \hyperref[fig:specific heat]{3c} depicts the results of QTM calculations which is seen to match very well with the experimental data to confirm once again that CuD is quite a good system to realise a spin-$\frac{1}{2}$ AfHc. 
\begin{figure*}
	\centering
	\includegraphics[width=1\linewidth]{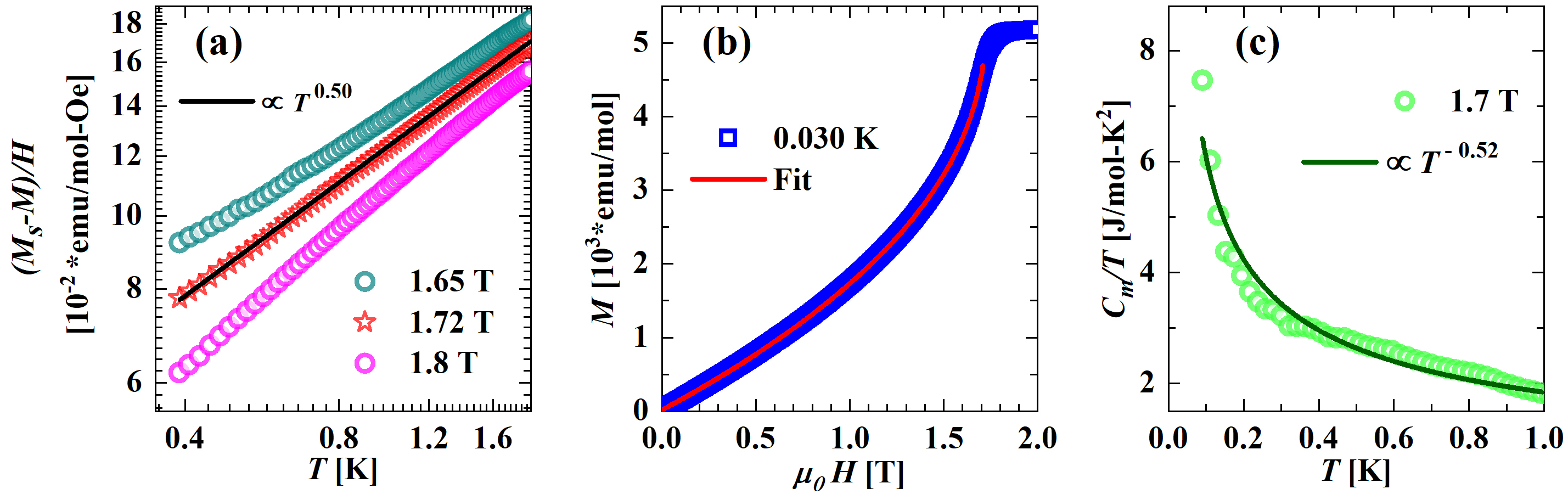}%
	\caption{(a) Log-log plot of $(M_s-M)/H$ as a function of temperature for fields close to the saturation field showing data for 1.65 T (olive green open circles), 1.72 T (red open stars) and 1.8 T (pink open circles). Black solid line is a straight line fit to the 1.72 T data yielding $\beta=0.5$ and $M_s=5216.26$ emu/mole (see text for details). (b) Blue open squares represent magnetisation as a function of applied field $\mu_0 H$ at a temperature of 0.030 K. Red solid line is a fit to the data (see text for details).  (c) Green open circles represent the temperature variation of $C_m(T)/T$ at an applied field of 1.7 T while dark green solid line is a power law fit, $T^{\alpha}$, to the data. The best fit yielded $\alpha = 0.52$.}
	\label{fig:critical exponents}
\end{figure*} 
From the free energy functional of equation \ref{eqn:free energy}, magnetisation $M$ in the saturation field quantum critical region is found to be independent of $\mu_0(H - H_s)$ (see SI section \hyperref[Supplementary]{7} for details) and varies as:  
\begin{equation}
	M_s - M = g\mu_B \zeta(0.5)(1-\sqrt{2})\bigg(\frac{k_B T}{4 \pi J}\bigg)^{\frac{1}{2}}
	\label{eqn:saturation M}
\end{equation} 
where $M_s$ is the saturation magnetisation.\\
The saturation field, $\mu_0 H_s$, was found experimentally by fitting $(M_s - M)/H$ vs. $T$ (obtained from Figs. \hyperref[fig:magnetisation]{2a} and \hyperref[fig:magnetisation]{2b}) to $T^{\beta}$ for different values of fields very close to $\mu_0 H_s$ as shown in Fig. \hyperref[fig:critical exponents]{4a}. The fits were made keeping both $M_s$ and $\beta$ as free parameters while having $J/k_B$ and $g$ to have fixed values of 1.23 K and  2.03 respectively (obtained from the magnetisation data analysis above). The best fit to data was obtained for an external applied field of 1.72 T that gave the value of saturation magnetisation $M_s$ as 5216.26 emu/mol close to the expected value of 5212 emu/mol (see Fig. \hyperref[fig:critical exponents]{4b}),  with $\beta = 0.50$, identical to the expected theoretical value of $\frac{1}{2}$ from equation \ref{eqn:saturation M}.\\
Magnetic field variation of magnetisation, $M$, obtained by integrating magnetic field dependent $\chi'(\mu_0H)$ measured in CuD at  $T = 0.030$ K is shown by blue squares in Fig. \hyperref[fig:critical exponents]{4b}. It can be seen that $M$ increases with field till $\sim 1.7 $ T ($\mu_0 H_s$) above which it saturates due to a transformation of the TLL state to a FP state. Red solid curve in Fig. \hyperref[fig:critical exponents]{4b} represents a fit the expression $1-M/M_s = 1.27(1-H/H_s)^{1/\delta}$ to the data between 0 T and 1.7 T considering $M_s = 5216.26 $ emu/mol yields critical field $\mu_0 H_s = 1.72$ T and $\delta = 2.2$, in agreement with the predicted value of critical exponent $\delta = 2$.\\
From the free energy expression \ref{eqn:free energy}, $C_m(T)/T$ should diverge with temperature at the saturation field $\mu_0 H_s$ since $C_m(T)/T$ is proportional to the density of states $D(\epsilon)$ which goes as $1/\sqrt{T}$ in one dimension. Such a divergence was indeed observed in CuD as shown in Fig. \hyperref[fig:critical exponents]{4c} where a fit of $C_m/T$ vs. $T$ data below 1 K (green open circles) to  $C_m/T = 0.22894N_a k_B^{3/2}(JT)^{-\alpha}$ (dark green solid line) keeping $J/k_B = 1.23$ K, gave a value of the exponent $\alpha$ as 0.52, very close to the expected value of 0.5 \cite{xiang}. The combination of power-law exponents extracted from thermodynamic measurements described above is $\alpha + \beta(1+\delta) = $ 2.12, where $\alpha = $ 0.52, $\beta = $ 0.5, and $\delta = $ 2.22, close to the theoretically expected universal scaling value of 2 \cite{he} confirming the excellent realisation of CuD as a spin-$\frac{1}{2}$ AfHc system.

\begin{figure*}
	\centering
	\includegraphics[width=1\linewidth]{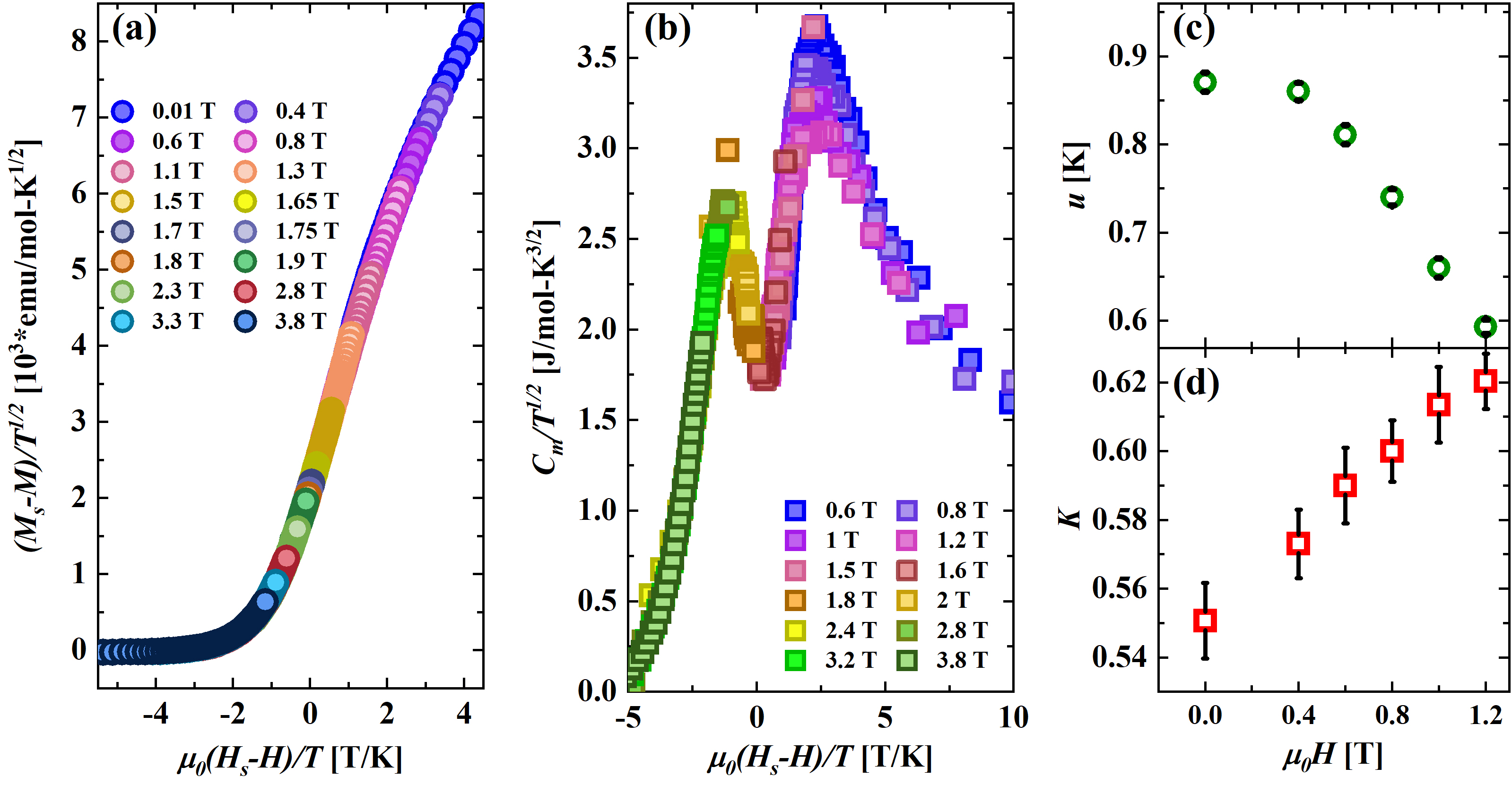}
	\caption{(a) Filled circles represent the scaling collapses of magnetization data as described in the text with $\mu_0 H_s=$ 1.72 T. (b) Filled squares represent the scaling collapses of specific heat data as described in the text where $\mu_0 H_s$ = 1.7 T. (c) Spinon velocity $u$ as a function of field calculated from $C_m(T)$ (see text for details). (d) The Luttinger parameter $K$ calculated from $\chi'(\mu_0 H)$ and spinon velocity $u$, is shown as a function of applied field.}
	\label{fig:Luttinger parameters}
\end{figure*}
At the saturation field quantum critical point $\mu_0 H_s$, the theory given by equation \ref{eqn:free energy} is invariant with respect to scaling transformations (see SI section \hyperref[Supplementary]{7} for details), resulting in magnetisation and specific heat exhibiting scaling. So a plot of ($M_s-M$)/$\sqrt{T/J}$ as a function of $\mu_0 (H_s - H)/T$ (with $\mu_0 H_s$ = 1.72 T) for different values of $\mu_0 H$ should result in a data collapse as shown in Fig. \hyperref[fig:Luttinger parameters]{5a}. A similar plot of $C/\sqrt{T}$ as a function of $\mu_0 (H_s - H)/T$ gives the data collapse for $\mu_0 H_s = 1.7$ T as seen in Fig. \hyperref[fig:Luttinger parameters]{5b}.\\
The TLL universality class is characterized by collective spinon excitations that are coherent and linearly dispersing at low energies \cite{hammar,cavazos}. Consequently, the molar specific heat of a TLL at low temperatures is given by \cite{giamarchi,bouillot2011}:
\begin{equation}
	C_{TLL} = N_A \frac{\pi k_B T}{6u}
	\label{eqn:C_TLL}
\end{equation}
where $N_A$ is the Avogadro constant and $u$ the spinon velocity. A straight line fit to the $T$-linear regime of the $C_m(T)$ data for $\mu_0 H < \mu_0 H_s$ (for 0 T data, the linear fit is shown as red line in the insets of Figs. \hyperref[fig:specific heat]{3a} and \hyperref[fig:specific heat]{3b}) yielded the spinon velocity in accordance with equation \ref{eqn:C_TLL}, as shown in Fig. \hyperref[fig:Luttinger parameters]{5c}. It is observed that $u$ decreases with increasing field, likely due to the increased presence of string-like domains of polarised spins within the system \cite{bera,he}.\\
Having obtained $u$, the other parameter of the field theory-Luttinger parameter $K$, was obtained from the relation between $\chi'(\mu_0H)$, $u$ and $K$ as \cite{giamarchi,schmidiger}:
\begin{equation}
	\chi'(\mu_0 H) = N_A\frac{(g\mu_B)^2}{2\pi k_B}\left(\frac{K}{u}\right)
	\label{eq:Luttinger_para}
\end{equation}
and utilising field dependent $\chi'(\mu_0 H)$ at 0.03 K and $u$ values obtained above. The calculated $K$ values are plotted as a function of $\mu_0 H$ in Fig. \hyperref[fig:Luttinger parameters]{5d}. It can be seen that $K$ is positive for all values of applied fields indicating that the interactions in TLL are repulsive in nature \cite{giamarchi,jeong2013,jeong2016,klanjvsek}. Furthermore, the value of $K$ is 0.55 for zero field, tantalisingly close to the expected value of 0.5 \cite{giamarchi} indicating that CuD realises the TLL state quite well.\\
To conclude, we show the first experimental phase diagram of a spin-$\frac{1}{2}$ antiferromagnetic Heisenberg chain by utilising Wilson ratio in single crystals of a new metal-organic compound \ce{C14H18CuN4O10} depicting Tomonaga-Luttinger liquid, field-induced quantum critical and fully polarized phases. The construction of the phase diagram over large range of fields and temperatures, and identification of different phases was enabled by the unique low energy scale of the exchange interactions in \ce{C14H18CuN4O10}. Wilson ratio was calculated using magnetization, magnetic susceptibility and specific heat measurements in \ce{C14H18CuN4O10}. Theoretical magnetisation and specific heat curves generated using quantum transfer matrix method were found to match the experimentally obtained data very well. Quantum critical phase boundaries were found to affect a large portion of the phase diagram. Consequently, quantum critical scaling obtained through field theoretical methods was shown to hold true over large areas of phase diagram through data collapse in magnetisation and specific heat. Finally, parameters of the Tomonaga-Luttinger liquid theory, namely, spinon velocity and Luttinger parameter were calculated was found to match with theoretical predictions very well. In order to probe the transformation of the spinon excitation spectrum (corresponding to the TLL state) to that of the magnon spectrum of the fully polarised state, we intend to perform inelastic neutron scattering measurements on CuD and construct a magnetic phase diagram. Such measurements would be facilitated by the availability of large sized single crystals of CuD as well as the low saturation field $\mu_0 H_s$ of $\sim$ 1.7 T easily accessible in various neutron facilities worldwide.
 
\matmethods{\subsection*{Synthesis}
CuD crystals were grown using liquid-liquid diffusion technique where 0.100 g (0.3 mmol) of potassium bis(oxalato)cuprate(II) dihydrate was dissolved in 12 ml of distilled water to make solution A. Similarly, 0.059 g (0.6 mmol) of 4-aminopyridine was dissolved in 12 ml of ethyl acetate to make the less dense solution B. Solution A was added to the bottom of a cylindrical glass vessel to create the bottom layer. The top layer was created using solution B that was added to the cylinder slowly creating a liquid/liquid boundary. Large sized single crystals of CuD formed at the bottom of the cylinder after 2 months. 
\subsection*{Measurements}
Single-crystal X-ray diffraction (SCXRD) measurements were done on a Bruker APEX II CCD diffractometer using graphite-monochromatized Mo-K$_\alpha$ radiation ($\lambda$ = 0.711 $\AA$) at a room temperature of 296(2) K. DC magnetic measurement were performed on a single-crystal of CuD having a mass of 14 mg in the temperature range of 1.8 $<$ T $<$ 150 K using a Quantum Design SQUID (Superconducting Quantum Interference Device) magnetometer (Model MPMS3). Magnetisation measurements in the temperature range of 0.38 $<$ T $<$ 2 K were done using a $^3$He insert attached to the MPMS3 (Model iHelium3). Magnetic susceptibility measurements were conducted using a custom-built ac susceptometer within a temperature range of 0.03 K to 0.8 K. An ac magnetic field with a magnitude of 1.2 Oe and a frequency of 471 Hz was generated by an ac current source (Stanford Research, CS 580), while a lock-in amplifier (Stanford Research, SR 830) was utilized to record the corresponding ac signal. The sample temperature was monitored by a calibrated \ce{RuO2} sensor positioned adjacent to the ac susceptometer. The magnetic susceptibility measured with the ac susceptometer can be considered equivalent to the under dc conditions, as the frequency of the ac magnetic field falls within the dc limits for the sample. The specific heat data were collected on a 0.85 mg single crystal in the temperature range 0.09 $<$ T $<$ 4 K using a Quantum Design PPMS equipped with a  $^3$He-$^4$He dilution refrigerator.
\subsection*{Theoretical models}
		The thermodynamic properties of the integrable spin-$\frac{1}{2}$ AfHc were obtained through a combination of the
quantum transfer matrix method and the Bethe Ansatz by Klumper \cite{klumper} by considering a linear
energy-momentum dispersion of the spinons having a velocity $v=\pi J$. A set of non-linear integral equations
were derived, the solutions of which determine the free energy. We have computed the derivatives of the free energy
equations (analytically), which have been used to obtain the equations for specific heat, magnetization and susceptibility. An efficient
iterative scheme utilizing fast Fourier transform has been implemented for solving these equations numerically over a wide range of temperatures
and fields (see SI section \hyperref[Supplementary]{6} for details of the derivations and the numerical implementation).\\
The effective spinless fermion field theory that describes the QCP near the saturation field starts with the Jordan-Wigner transformation for the spin chain 
to arrive at a spinless fermionic form. The Heisenberg
model 
$H = J \sum_{i} \mathbf{S}_i \cdot \mathbf{S}_{i+1} + \sum_i g \mu_B \mathbf{H} \cdot \mathbf{S}_i$
with $\mathbf{H}  = H\hat{z}$ gives the polarized state
$|\ldots \downarrow \downarrow \downarrow \downarrow \ldots \rangle$ above the saturation field
$H_s$. Taking it as the fermionic vaccum, the mapping is
$S^+_i = \prod_{k < i} (-1)^{n_k} c^\dagger_i$,
$S^-_i = \prod_{k < i} (-1)^{n_k} c_i$ and $S^z_i = c^\dagger_i c_i - \frac{1}{2}
= n_i - \frac{1}{2}$ where $i \in \mathbf{Z}$ tracks sites on the spin chain.
This leads to the fermionic Hamiltonian,
$ \sum_{i} \left( \frac{J}{2} ( c^\dagger_i c_{i+1} + c^\dagger_{i+1} c_i )
+ J n_i n_{i+1} \right)
+ \sum_i \left( (J_H - J) n_i + (J-2J_H)/4 \right)$
where $J_H = g \mu_B H$.
A continuum field $\psi(x)$ is built out of the lattice fermion degrees of
freedom by setting $\psi(x_i) = (-1)^i c_i/\sqrt{a}$ where $a$ is the
lattice parameter between two neighbouring spins. Doing a
gradient expansion $\psi(x_{i+1}) = \psi(x_i + a) \sim \psi_(x_i)
+ a \partial_x \psi(x) + \frac{a^2}{2} \partial^2_x \psi(x) \ldots$, one arrives at 
the continuum theory
\begin{equation}
	\int dx \left(Ja^2 \; \psi^\dagger(x) \partial^2_x \psi(x) 
	- (2J - J_H) \; \psi^\dagger(x) \psi(x) \right) 
	\label{eq:free_fermion_field_theory}
\end{equation}
for the quadratic terms, and
\begin{equation}
	\int dx \left(-J a^3 \; \psi^\dagger(x) \partial_x \psi^\dagger(x) \psi(x) \partial_x \psi(x) \right) 
	\label{eq:interaction_term_field_theory}
\end{equation}
for the interaction term. A contact term 
$\propto \int dx \left(\psi^\dagger(x) \psi(x)\right)^2$ is absent due to the spinless
nature of the fermions.
		
}
	
\showmatmethods{}
	
		\subsection*{Supplementary Material}
	\label{Supplementary}
	Supplementary material is available at PNAS Nexus online.
	
\subsection*{Data, Materials, and Software Availability} All  study data  are included  in the article and/or SI Appendix.

\acknow{D.J-N acknowledges financial support from SERB, DST, Govt. of India (Grant No. CRG/2021/001262). FIST facility at IISER Thiruvanthapuram is acknowledged for providing the cryogenic environment (Grant No. SR/FST/PS-II/2018/54 [C]). N.S.V. thanks the Department of Science and Technology, Government of India, for funding through the national supercomputing mission grant DST/NSM/R$\&$D$\_$HPC$\_$Applications/2021/26. S.P. acknowledges financial support from SERB, DST, Govt. of India (Grant No. MTR/2022/000386). S.L. thanks the SERB, Govt. of India for funding through MATRICS grant MTR/2021/000141 and Core Research Grant CRG/2021/000852. A portion of this work was performed at the National High Magnetic Field Laboratory, which is supported by National Science Foundation Cooperative Agreement No. DMR-2128556 and the State of Florida.}
	
\showacknow{}

\balance	
\end{document}


\maketitle


\section{Crystal structure}

\begin{figure}
	\centering
	\includegraphics[width=1\textwidth]{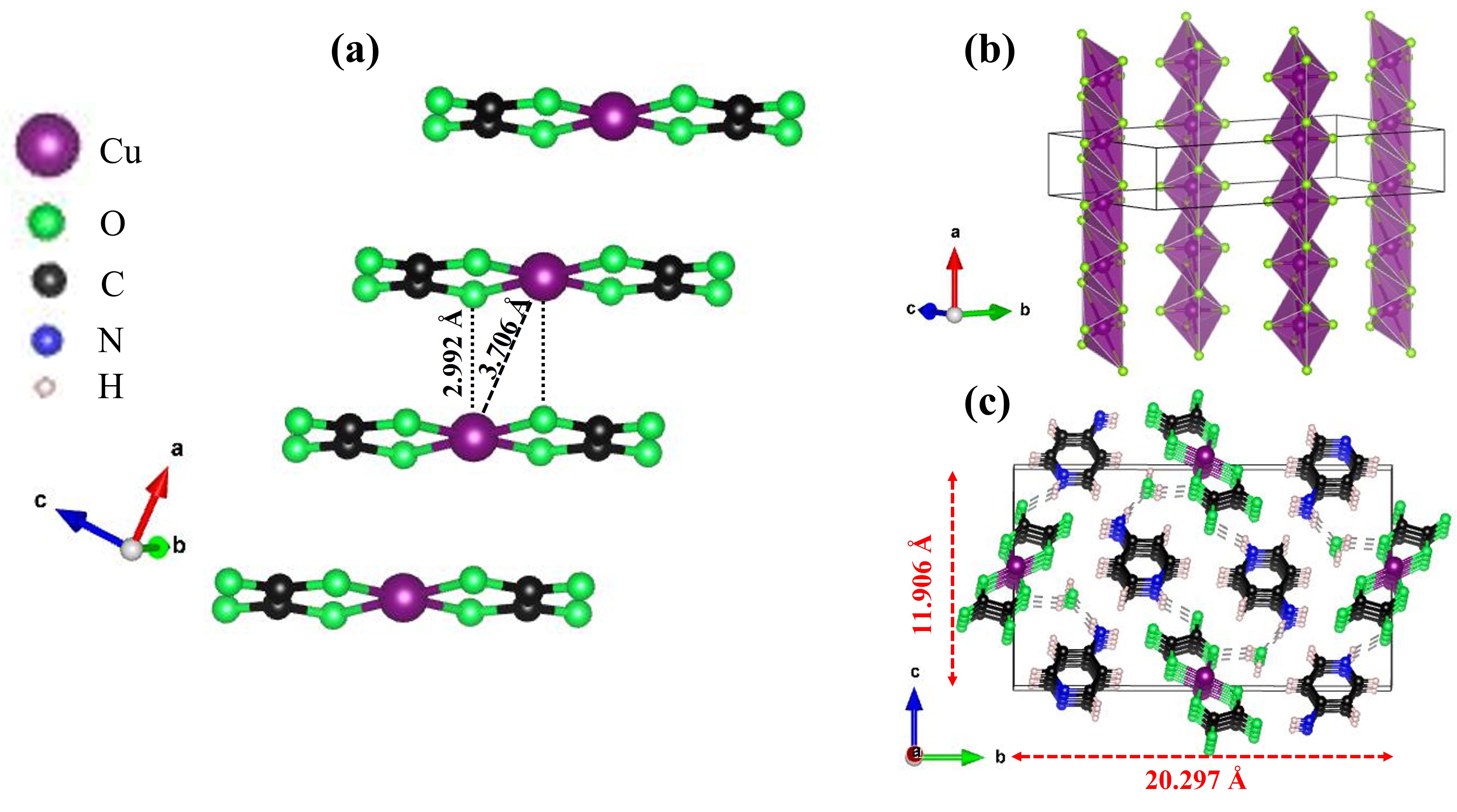}
	\caption{(a) Chain structure formation in \ce{C14H18CuN4O10} along the $a$-axis through stacking of \ce{[Cu(C2O4)2]^{-2}} units. (b) Infinite chains of \ce{Cu^{2+}O6} octahedra sharing common apical oxygen atoms. (c) $bc$ plane of \ce{C14H18CuN4O10} with Cu--Cu distances of 20.297 $\AA$ along the $b$-axis and 11.906 $\AA$ along the $c$-axis.}
	\label{fig:crystal structure}
\end{figure}

In Fig. \ref{fig:crystal structure}, the structure of \ce{C14H18CuN4O10} features corner-sharing \ce{CuO6} octahedra that form Cu–Cu chains along the $a$-axis. These chains are well isolated in the $b$ and $c$ directions.

\section{Magnetisation}
\begin{figure}
	\centering
	\includegraphics[width=1\textwidth]{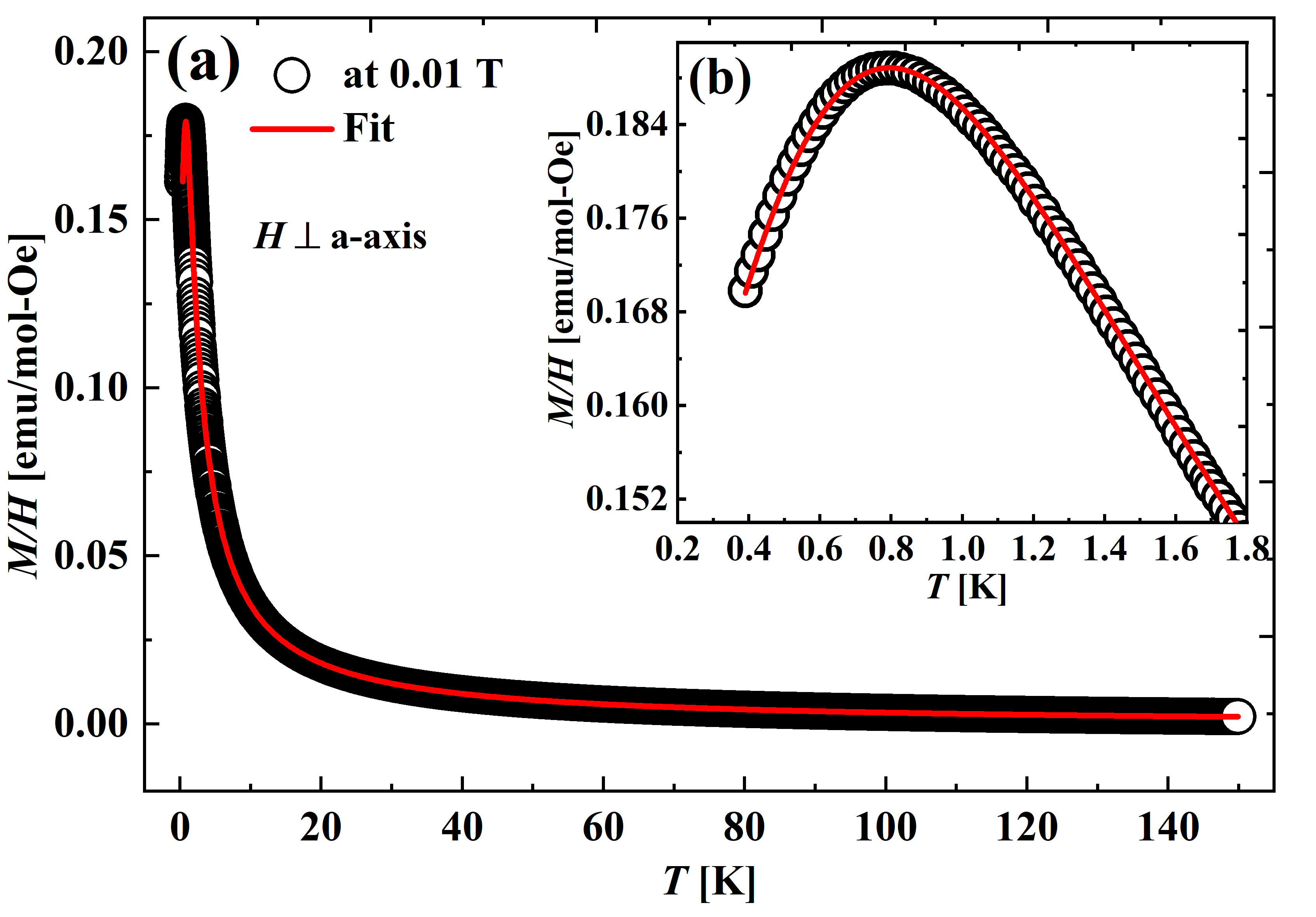}
	\caption{(a) Temperature variation of magnetisation at an applied field of 0.01 T in \ce{C14H18CuN4O10}. Open circles represent the data points, while the red solid curve is a fit to the uniform spin-$\frac{1}{2}$ AfHc model. (b) M-T data shown in the lower temperature range of 0.4 K to 1.8 K to show the peak in (a) on an expanded scale.}
	\label{fig:magnetisation_B-F}
\end{figure}
The exchange coupling constant $J/k_B$ between \ce{Cu^{2+}} ions was determined by fitting the temperature dependent magnetisation data to the following equation: 

\begin{equation}
M(T)/H = \chi_0 + \frac{C_{para}}{T} + \chi_{spin}(T)
\label{eq:B-F}
\end{equation}

where $\chi_0$ represents the diamagnetic contribution arising due to closed shell structure of the atoms; $\frac{C_{para}}{T}$ the paramagnetic contribution arising due to impurities, and $\chi_{spin}(T)$ the spin susceptibility.
The high-temperature series expansion equation of magnetic susceptibility  of a spin-$\frac{1}{2}$ uniform AfHc is given by \cite{bonner}:

\begin{equation}
\chi_{spin}(T) = \frac{N\mu_B^2 g^2}{k_B T}\frac{0.25+0.0775x+0.0752x^2}{1+0.993x+0.1721x^2+0.7578x^3}
\end{equation}

where the symbols have their usual meanings and $x = J/k_BT$. In Fig. \ref{fig:magnetisation_B-F}, the best fit of eq. \ref{eq:B-F} to the experimental data is shown as a red solid curve. The fit parameters obtained were $\chi_0 = -4.16\times10^{-4}$, paramagnetic impurity concentration $C_{para} = 0.33$, intra-coupling constant $J/k_B = 1.23$ K and Land\'{e} factor $g$ = 2.03.

\section{Specific heat and Debye-Einstien model}

\begin{figure}
	\centering
	\includegraphics[width=1\textwidth]{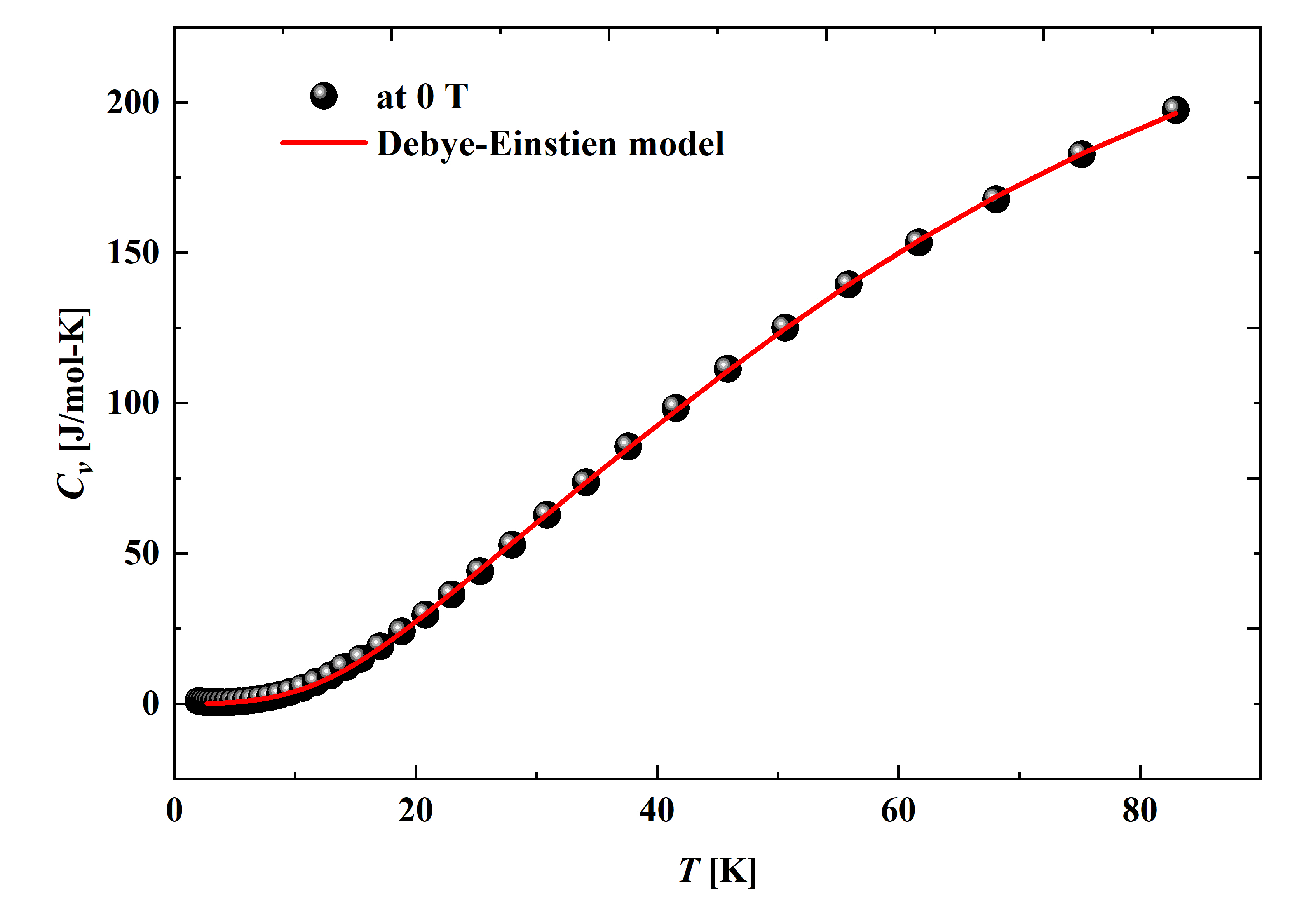}
	\caption{Temperature variation of specific heat in \ce{C14H18CuN4O10} in the temperature range of 2 K to 80 K. Solid spheres represent the zero field data while the red solid curve is a fit to the Debye-Einstein model.} 
	\label{fig:specific_heat_D-E}
\end{figure}

In order to evaluate the magnetic component of the specific heat, $C_m(T)$, the phonon contribution to the specfic heat, $C_p$, needs to be subtracted from the total specific heat $C_v$ obtained experimentally and shown in Fig. \ref{fig:specific_heat_D-E}. This was done by fitting $C_v$ to the Debye-Einstein model using a linear combination of one Debye and two Einstein terms as shown in the equation below \cite{kittel}:

\begin{equation}
C_p(T) = m C_D(\theta_D,T)+\sum_{i=1}^{2}t_i C_{E_{i}}(\theta_{E_{i}},T)
\label{eq:debye-einstien}
\end{equation}

where $m$ and $t_i$ represent the weightage of the Debye and Einstien terms respectively. $C_D$ and $C_E$ are the Debye and Einstein contributions to the $C_p$, as given by the following equations:

\begin{equation}
C_D(T) = 9nR \left(\frac{T}{\theta_D}\right) \int_{0}^{\frac{\theta_D}{T}} \frac{X^4 e^x}{(e^x - 1 )^2} dx
\label{eq:debye}
\end{equation}

\begin{equation}
C_E(T) = 3nR\left(\frac{\theta_{E}}{T}\right)^2 \frac{e^{\frac{\theta_E}{T}}}{(e^{\frac{\theta_E}{T}}-1)^2}
\label{eq:einstien}
\end{equation}

where $n$, $R$, $\theta_D$ and $\theta_E$ are the number of atoms per formula unit, universal gas constant, Debye temperature and Einstein temperature, respectively. The fitted curve is shown in Fig. \ref{fig:specific_heat_D-E} as a red solid curve. The obtained fitting parameters are $\theta_D= 136.96$ K, $\theta_{E_{1}}= 247.97$ K, $\theta_{E_{2}} = 4548.13$ K, $m=0.115$, $t_1 = 0.132$, and $t_2 = 0.782$.

\section{Coefficient of magnetic contribution to specific heat}
\begin{figure}
	\centering
	\includegraphics[width=1\textwidth]{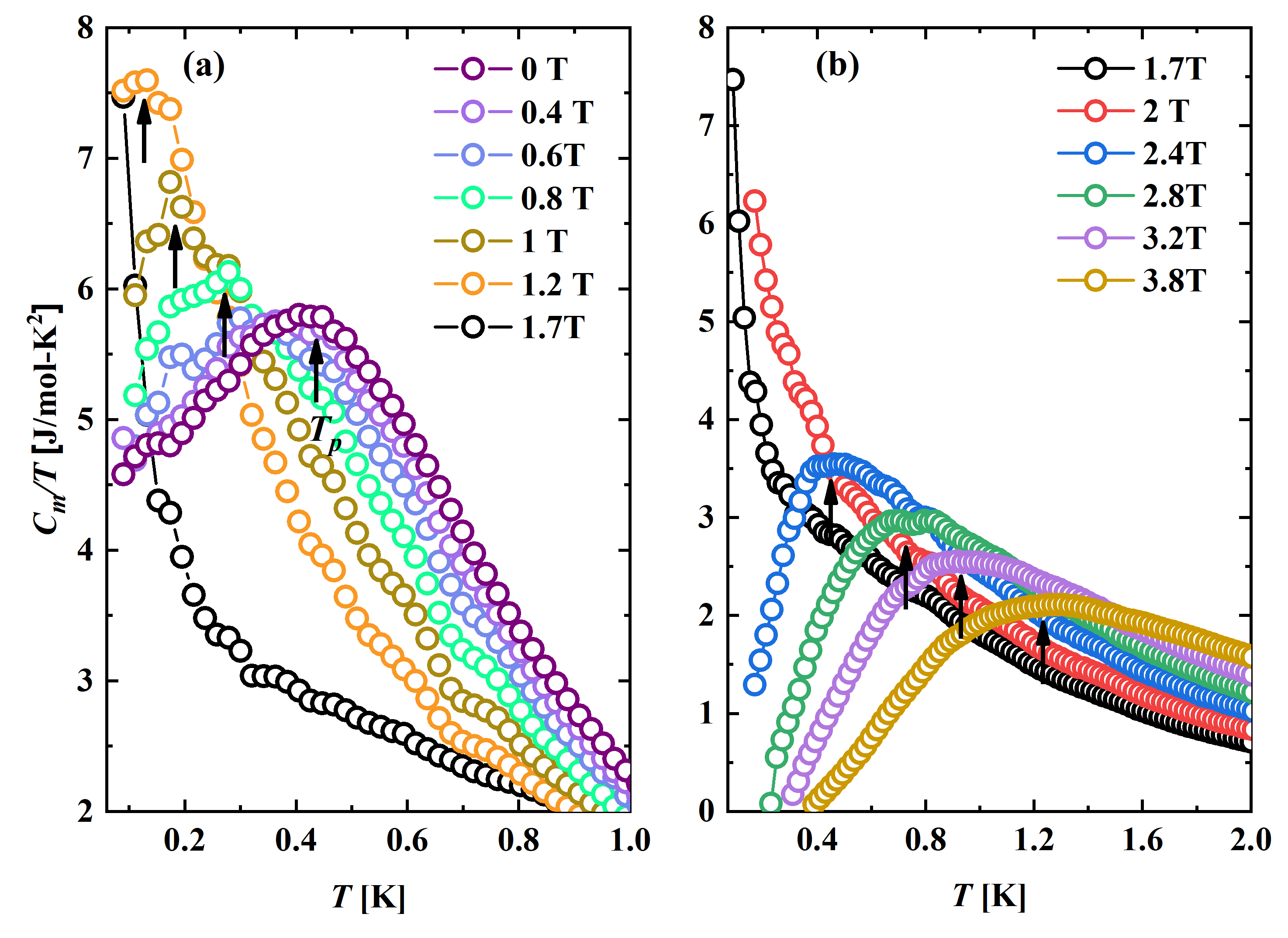}
	\caption{The magnetic contribution to specific heat divided by temperature in \ce{C14H18CuN4O10} is plotted as a function of temperature for various field values until (a) saturation field $\mu_0 H_s$ (b) above $\mu_0 H_s$.} 
	\label{fig:specific_heat_by_Temp}
\end{figure} 
In Fig. \ref{fig:specific_heat_by_Temp}, $C_m(T)/T$ is plotted as a function of temperature for various magnetic field values. The peak temperature ($T_p$) is marked by an arrow and shown as open circles in Fig. 1 of the main article. $T_p$ is found to decrease steadily with an increase in the applied magnetic field until 1.2 T above which it falls below the lowest measurable temperature of 0.1 K. At fields near $\mu_0 H_s$, $C_m(T)/T$ exhibits divergence as $T\rightarrow0$ K, indicating quantum criticality \cite{klumper1998,xiang}. For fields above $\mu_0 H_s$, $T_p$ appears above 2 T and moves to higher temperatures with further increase in field.

\section{Wilson ratio}

\begin{figure}
	\centering
	\includegraphics[width=1\textwidth]{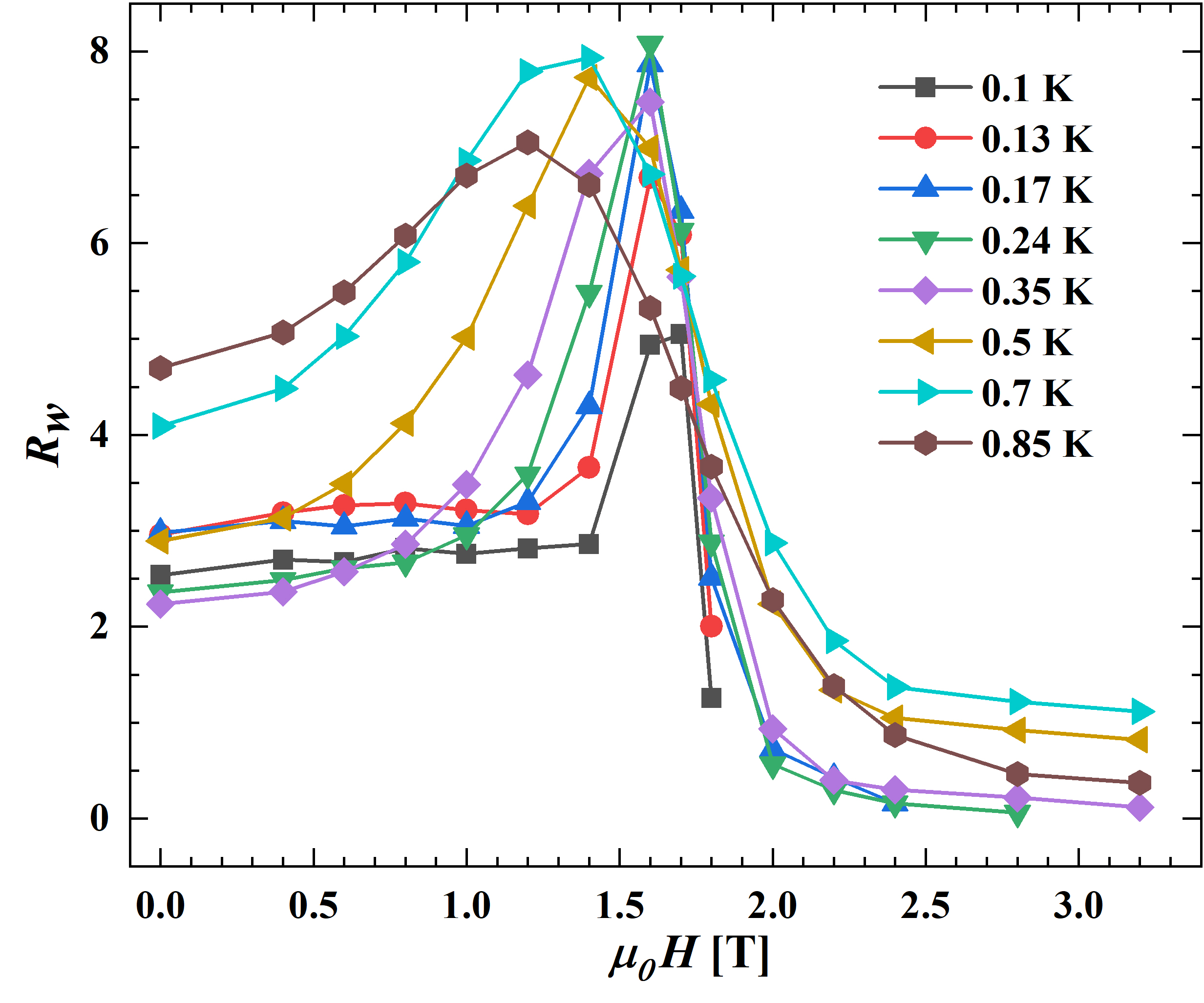}
	\caption{Wilson ratio in \ce{C14H18CuN4O10} is plotted as a function of magnetic field for different values of temperature.} 
	\label{fig:wilson_ratio}
\end{figure}  

The Wilson ratio, $R_w$, is estimated by using the formula:
\begin{equation}
R_w = \frac{4}{3}\bigg(\frac{\pi k_B}{\mu_B g}\bigg)^2\frac{\chi'}{C_m / T}
\label{eq:wilson_ratio}
\end{equation}
where the symbols have their usual meanings. $\chi$ represents the magnetic susceptibility while $C_m/T$ denotes the coefficient of the magnetic contribution to specific heat. In Fig. \ref{fig:wilson_ratio}, the estimated Wilson ratio is plotted as a function of field for different values of temperature. The fields corresponding to the maximum value of $R_w$ are marked as open stars in Fig. 1 of the main article.

\section{XXX Model}

\subsection{Introduction}
Using the quantum transfer matrix approach and the Bethe ansatz, a set of equations to compute free energy for the uniform and the anisotropic one-dimensional spin-$\frac{1}{2}$ Heisenberg model were derived by Klumper\cite{klumper1998,klumper2008integrability,klumper1993thermodynamics}. The free energy per lattice site of the spin-$\frac{1}{2}$, uniform, 1-d AfHc, with exchange coupling $J$, is expressed in terms of auxiliary functions, $\mathfrak{U}(x)$ and $\Bar{\mathfrak{U}}(x)$ as~\cite{klumper1998}:

\[ \beta f = \beta e_{0} - \frac{1}{2\pi}\int_{-\infty}^{+\infty}\frac{log[\mathfrak{U}\Bar{\mathfrak{U}}(x)]}{cosh(x)}dx\,.\]
The auxiliary functions are, in turn, given by the following set of non-linear integral equations:
\[log\mathfrak{a}(x) = \frac{-\nu\beta}{coshx} + \phi + \int_{-\infty}^{+\infty}[K(x-y)log\mathfrak{U}(y) - K(x - y -i\pi + i\epsilon)log\Bar{\mathfrak{U}}(y)]dy\]

\[log\Bar{\mathfrak{a}}(x) = \frac{-\nu\beta}{coshx} - \phi + \int_{-\infty}^{+\infty}[K(x-y)log\Bar{\mathfrak{U}}(y) - K(x - y +i\pi - i\epsilon)log\mathfrak{U}(y)]dy\]
where $\mathfrak{U}(x) =1+ {\mathfrak{a}}$, and $\Bar{\mathfrak{U}}(x) =1+ \Bar{\mathfrak{a}}$; $v = \pi J$; $\phi = \frac{\beta}{2}h$, with $h=g\mu_B H$ and $H$ is the applied magnetic field. $K(x)$ is the integration kernel given by
\[K_1(x) = \frac{1}{\pi}\int_{0}^{\infty}\frac{e^{-\pi k}}{\cosh(\pi k)}\cos(2kx)dk\]

The Integral is computed and the plot of the function is shown below

\begin{figure}[H]
	\centering
	\includegraphics[width =1 \textwidth]{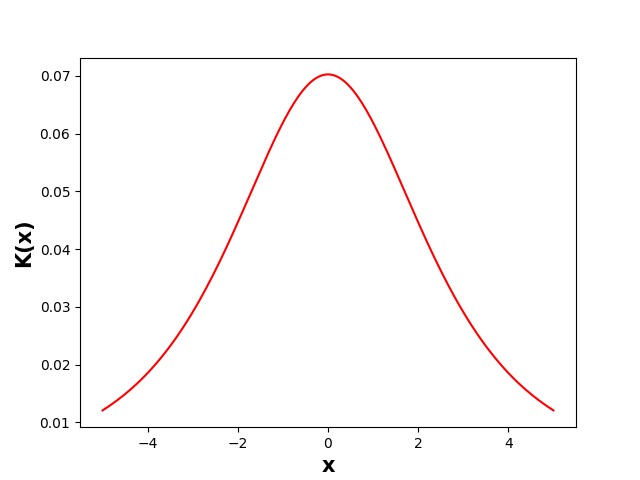}
	\caption{Plot of K(x)}
	\label{fig:my_label}
\end{figure}


The equation for the Integration Kernel, K(x-i$\pi$+i$\epsilon$) is given below,

\[K_2(x) =K_1(x-i\pi+i\epsilon) = \frac{1}{2\pi}\int_{0}^{\infty}\frac{e^{\frac{-\pi k}{2}}}{\cosh(\frac{\pi k}{2})}\cos(kx -i\pi +i\epsilon)dk\]

The Integral is computed and the plot of the function is shown below

\begin{figure}[t!]
	\centering
	\includegraphics[width=1\linewidth]{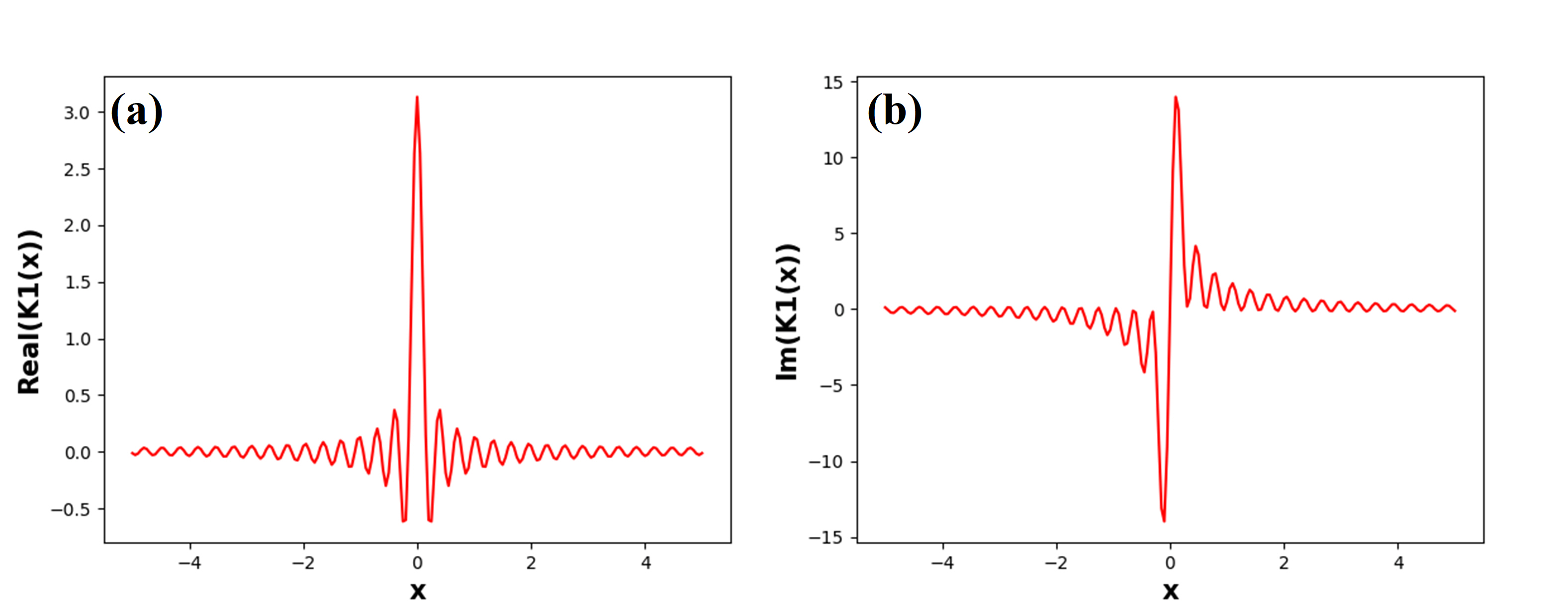}
	\phantomsubfigure{1a}%
	\phantomsubfigure{1b}%

	\caption{(a) Re(K$_1$(x)). (b) Im(K$_1$(x))  } 
	\label{fig:}
\end{figure}


The equation for the Integration Kernel, K(x+i$\pi$-i$\epsilon$) is given below,

\[K_3(x) = K_1(x+i\pi-i\epsilon)=\frac{1}{2\pi}\int_{0}^{\infty}\frac{e^{\frac{-\pi k}{2}}}{\cosh(\frac{\pi k}{2})}\cos(kx+i\pi-i\epsilon)dk\]

For $\epsilon \rightarrow 0 $

\begin{figure}[t!]
	\centering
	\includegraphics[width=1\linewidth]{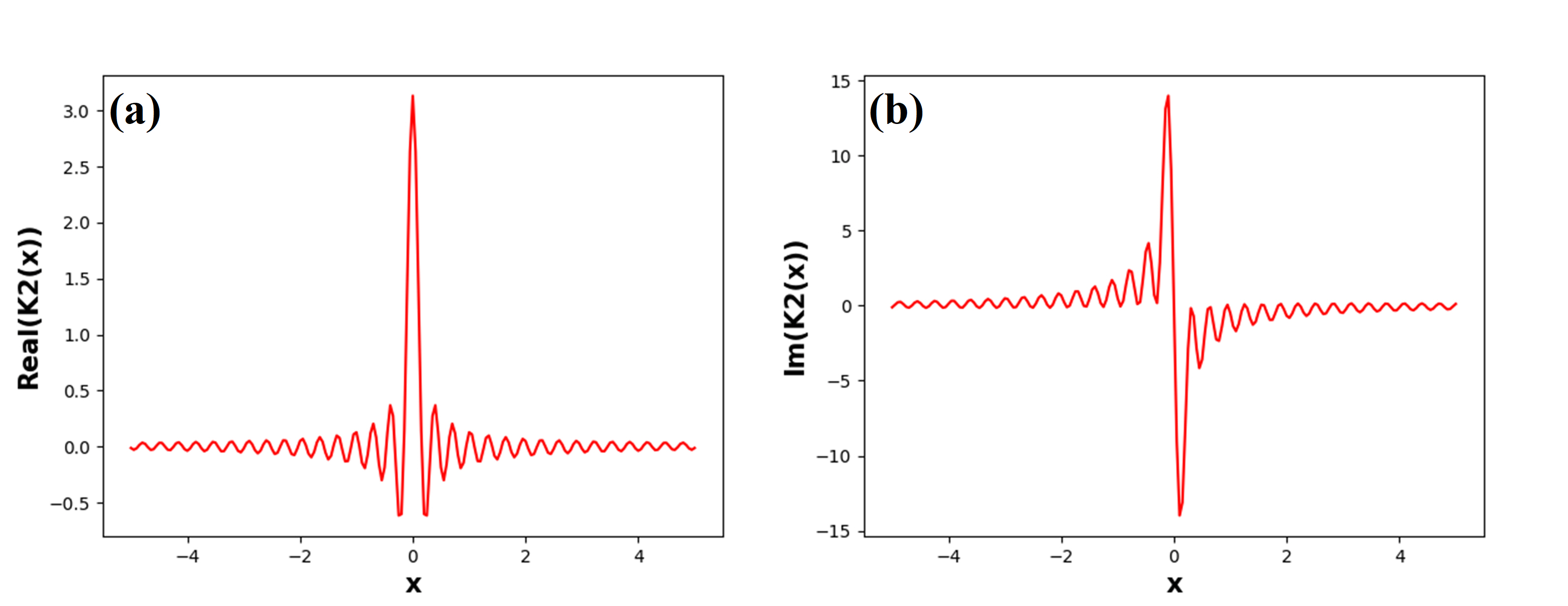}
	\phantomsubfigure{1a}%
	\phantomsubfigure{1b}%

	\caption{(a) Re(K$_2$(x)). (b) Im(K$_2$(x))  } 
	\label{fig:}
\end{figure}

\subsection{Numerical implementation of XXX Model}

\subsubsection{Integration Kernels}

\[K_1(x) = \frac{1}{2\pi}\int_{0}^{\infty}\frac{e^{\frac{-\pi k}{2}}}{\cosh(\frac{\pi k}{2})}\cos(kx)dk\]

\[K_2(x) = K_1(x-i\pi+i\epsilon) = \frac{1}{2\pi}\int_{0}^{\infty}\frac{e^{\frac{-\pi k}{2}}}{\cosh(\frac{\pi k}{2})}\cos(kx -i\pi +i\epsilon)dk\]

\[K_2(x) = \frac{1}{2\pi}\int_{0}^{\infty}\frac{e^{\frac{-\pi k}{2}}}{\cosh(\frac{\pi k}{2})}[coskxcosh(k(\pi-\epsilon)) + isinkxsinh(k(\pi - \epsilon))]dk\]

\[K_3(x) = [K_2(x)]^*\]

To avoid numerical overflow we do the following,

\[\frac{e^{\frac{-\pi k}{2}}}{\cosh(\frac{\pi k}{2})} = \frac{2e^{\frac{-\pi k}{2}}}{e^{\frac{\pi k}{2}}+e^{\frac{-\pi k}{2}}} =  \frac{2e^{{-\pi k}{}}}{1+e^{{-\pi k}{}}}\]

\[\frac{e^{\frac{-\pi k}{2}}}{\cosh(\frac{\pi k}{2})}cosh(k(\pi -\epsilon)) = \frac{e^{{-\pi k}{}}}{1+e^{{-\pi k}{}}}\left(e^{k(\pi-\epsilon)} + e^{-k(\pi-\epsilon)}\right) = \frac{e^{-k\epsilon}+ e^{-k(2\pi - \epsilon)}}{1 + e^{-\pi k}} \]

\[\frac{e^{\frac{-\pi k}{2}}}{\cosh(\frac{\pi k}{2})}sinh(k(\pi -\epsilon)) = \frac{e^{{-\pi k}{}}}{1+e^{{-\pi k}{}}}\left(e^{k(\pi-\epsilon)} - e^{-k(\pi-\epsilon)}\right) = \frac{e^{-k\epsilon} - e^{-k(2\pi - \epsilon)}}{1 + e^{-\pi k}} \]

At $\epsilon \rightarrow 0$,

\[\frac{e^{\frac{-\pi k}{2}}}{\cosh(\frac{\pi k}{2})}cosh(k(\pi -\epsilon)) = \frac{1 + e^{-2\pi k}}{1 + e^{-\pi k}}\]

\[\frac{e^{\frac{-\pi k}{2}}}{\cosh(\frac{\pi k}{2})}sinh(k(\pi -\epsilon)) = \frac{1 - e^{-2\pi k}}{1 + e^{-\pi k}}\]

\subsubsection{Derivatives Implementation for specific heat computation}

Let us define the convolution function,$\mathfrak{C}(g(x),h(x))$ as shown below
\[\mathfrak{C}(g(x),h(x)) = \int_{-\infty}^{\infty}g(\tau)h(x-\tau)d\tau\]

Using the above definition the equations are

\[log\mathfrak{a}(x) = \frac{-\nu\beta}{coshx} + \phi + \mathfrak{C}(K(x),log\mathfrak{U}(x)) - \mathfrak{C}(K(x-i\pi +i\epsilon),log\Bar{\mathfrak{U}}(x))\]

\[log\mathfrak{a}(x) = \frac{-\nu\beta}{coshx} + \phi + \mathfrak{C}(K(x),log\Bar{\mathfrak{U}}(x)) - \mathfrak{C}(K(x+i\pi -i\epsilon),log{\mathfrak{U}}(x))\]

Multiplying by $k_bT/J = \Bar{T}$

\[\Bar{T}log\mathfrak{a}(x) = \frac{-\pi}{coshx} + \frac{h}{2} + \mathfrak{C}(\Bar{T}log\mathfrak{U}(x),K_1(x)) - \mathfrak{C}(\Bar{T}log\mathfrak{\Bar{U}}(x),K_2(x))\]

\[\Bar{T}log\mathfrak{a}(x) = \frac{-\pi}{coshx} - \frac{h}{2} + \mathfrak{C}(\Bar{T}log\mathfrak{\Bar{U}}(x),K_1(x)) - \mathfrak{C}(\Bar{T}log\mathfrak{U}(x),K_3(x))\]

Let us define $\Bar{T}log\mathfrak{U}(x) = L_u$ \qquad $\Bar{T}log\mathfrak{\Bar{U}}(x) = L_{\Bar{u}}$ \qquad $\Bar{T}log\mathfrak{a}(x) = L_a$  \qquad  $\Bar{T}log\mathfrak{\Bar{a}}(x) = L_{\Bar{a}}$

Then the equations become,

\[\boxed{ L_a = \frac{-\pi}{coshx} + \frac{h}{2} + \mathfrak{C}(L_u,K_1) - \mathfrak{C}(L_{\Bar{u}},K_2)}\]

\[\boxed{L_{\Bar{a}} = \frac{-\pi}{coshx} - \frac{h}{2} + \mathfrak{C}(L_{\Bar{u}},K_1) - \mathfrak{C}(L_u,K_3)}\]

We know that,

\[\mathfrak{U} = 1 + \mathfrak{a}\]

\[\Bar{T}log\mathfrak{U} = \Bar{T}log(1+\mathfrak{a})\]

\[\boxed{L_u = \Bar{T}log\left(1 + exp\left(\frac{L_a}{T}\right)\right)}\] 

Similarly,

\[\boxed{L_{\Bar{u}} = \Bar{T}log\left(1 + exp\left(\frac{L_{\Bar{a}}}{T}\right)\right)} \]

\subsubsection{First Derivatives}

Taking the first derivatives of the boxed equations we get,

\[\frac{dL_a}{d\Bar{T}} = \mathfrak{C}(\frac{dL_u}{d\Bar{T}},K_1) - \mathfrak{C}(\frac{dL_{\Bar{u}}}{d\Bar{T}},K_2)\]

\[\frac{dL_{\Bar{a}}}{d\Bar{T}} = \mathfrak{C}(\frac{dL_{\Bar{u}}}{d\Bar{T}},K_1) - \mathfrak{C}(\frac{dL_u}{d\Bar{T}},K_3)\]

\[T\frac{dL_a}{d\Bar{T}} = L_u + exp\left(\frac{L_a - L_u}{\Bar{T}}\right)\left(\Bar{T}\frac{dL_a}{dT} - L_a\right) \]

\[T\frac{dL_{\Bar{a}}}{d\Bar{T}} = L_{\Bar{u}} + exp\left(\frac{L_{\Bar{a}} - L_{\Bar{u}}}{\Bar{T}}\right)\left(\Bar{T}\frac{dL_{\Bar{a}}}{dT} - L_{\Bar{a}} \right) \]

\subsubsection{Second Derivatives}

Taking the second derivatives of the above equations,

\[\frac{d^2L_a}{d\Bar{T}^2} = \mathfrak{C}(\frac{d^2L_u}{d\Bar{T}^2},K_1) - \mathfrak{C}(\frac{dL_{\Bar{u}}^2}{d\Bar{T}^2},K_2)\]

\[\frac{d^2L_{\Bar{a}}}{d\Bar{T}^2}  = \mathfrak{C}(\frac{dL_{\Bar{u}}^2}{d\Bar{T}^2},K_1) - \mathfrak{C}(\frac{dL_u^2}{d\Bar{T}^2},K_3)\]
\[\Bar{T}^3\frac{d^2L_a}{d\Bar{T}^2}  =\left(\Bar{T}\frac{dL_u}{d\Bar{T}} - L_u \right)\left[\frac{\Bar{T}^3\frac{d^2L_a}{dT^2}}{T\frac{dL_a}{dT} - L_a} + T\frac{dL_a}{dT} - L_a - T\frac{dL_u}{dT} - L_u\right] \]

\[\Bar{T}^3\frac{d^2L_{\Bar{a}}}{d\Bar{T}^2} =\left(\Bar{T}\frac{dL_{\Bar{u}}}{d\Bar{T}} - L_{\Bar{u}} \right)\left[\frac{\Bar{T}^3\frac{d^2L_{\Bar{a}}}{dT^2}}{\left(T\frac{dL_{\Bar{a}}}{dT} - L_{\Bar{a}}\right)} + \left(T\frac{dL_{\Bar{a}}}{dT} - L_{\Bar{a}}\right) - \left(T\frac{dL_{\Bar{u}}}{dT} - L_{\Bar{u}}\right)\right] \]

These equations are iteratively implemented to find $L_a, \frac{dL_a}{dT}, \frac{d^2L_a}{dT^2}$ and $L_{\Bar{a}}, \frac{dL_{\Bar{a}}}{dT}, \frac{d^2L_{\Bar{a}}}{dT^2}$ and $L_a, \frac{dL_u}{dT}, \frac{d^2L_u}{dT^2}$ and $L_{\Bar{u}}, \frac{dL_{\Bar{u}}}{dT}, \frac{d^2L_{\Bar{u}}}{dT^2}$.


\subsubsection{Specific heat}

We know that the equation for free energy is given by the equation,

\[ \beta f = \beta e_{0} - \frac{1}{2\pi}\int_{-\infty}^{+\infty}\frac{log[\mathfrak{U}\Bar{\mathfrak{U}}(x)]}{cosh(x)}dx\]

\[f/J = e_0 - \frac{T'}{2\pi}\int_{-\infty}^{+\infty}\frac{log[\mathfrak{U}\Bar{\mathfrak{U}}(x)]}{cosh(x)}dx\]

\[f/J = e_0 - \frac{1}{2\pi}\int_{-\infty}^{+\infty}\frac{dx}{coshx}\left[L_u + L_{\Bar{u}}\right]\]

\[\boxed{C_v = \frac{1}{2\pi}\int_{-\infty}^{+\infty}\frac{dx}{coshx}\left[\Bar{T}\frac{d^2L_u}{dT^2} + \Bar{T}\frac{d^2L_{\Bar{u}}}{dT^2}\right]}\]

\subsubsection{Derivatives implementation for susceptibility}

The first set of equations determining $C_v$ are the ones for free energy for a given (T,h). These equations remain the same. Only the derivatives change. So we will focus only on the derivatives.


\subsubsection{First derivatives}

\[{\frac{dL_a}{dh} = \frac{1}{2} + \mathfrak{C}(\frac{dL_u}{dh},K_1) - \mathfrak{C}(\frac{dL_{\Bar{u}}}{dh},K_2)}\]

\[{\frac{dL_{\Bar{a}}}{dh} = -\frac{1}{2} + \mathfrak{C}(\frac{dL_{\Bar{u}}}{d\Bar{T}},K_1) - \mathfrak{C}(\frac{dL_u}{dh},K_3)}\]

\[{T\frac{dL_u}{dh} =  exp\left(\frac{L_a - L_u}{\Bar{T}}\right)\frac{dL_a}{dT}}\]

\[{T\frac{dL_{\Bar{u}}}{d\Bar{T}} =  exp\left(\frac{(L_{\Bar{a}} - L_{\Bar{u}}}{\Bar{T}}\right)\frac{dL_{\Bar{a}}}{dT} }\]


\subsubsection{Second derivatives}

\[\frac{d^2L_a}{dh^2} = \mathfrak{C}(\frac{d^2L_u}{dh^2},K_1) - \mathfrak{C}(\frac{dL_{\Bar{u}}^2}{dh^2},K_2)\]

\[\frac{d^2L_{\Bar{a}}}{dh^2}  = \mathfrak{C}(\frac{dL_{\Bar{u}}^2}{dh^2},K_1) - \mathfrak{C}(\frac{dL_u^2}{dh^2},K_3)\]

\[\frac{d^2L_u}{dh^2} =  exp\left(\frac{L_a - L_u}{\Bar{T}}\right)\left[\frac{d^2L_a}{dh^2} + \frac{1}{T}\frac{d^2L_a}{dh^2}\left(\frac{d^2L_a}{dh^2} - \frac{d^2L_u}{dh^2} \right)\right]\]

\[\frac{d^2L_{\Bar{u}}}{dh^2} = exp\left(\frac{L_{\Bar{a}} - L_{\Bar{u}}}{\Bar{T}}\right)\left[\frac{d^2L_{\Bar{a}}}{dh^2} + \frac{1}{T}\frac{d^2L_{\Bar{a}}}{dh^2}\left(\frac{d^2L_{\Bar{a}}}{dh^2} - \frac{d^2L_{\Bar{u}}}{dh^2} \right)\right]\]

\subsubsection{Susceptibility}
Similar to the specific heat expression we can derive the susceptibility expression as the following, 

\[\boxed{\chi = \frac{1}{2\pi}\int_{-\infty}^{+\infty}\frac{dx}{coshx}\left[\Bar{T}\frac{d^2L_u}{dh^2} + \Bar{T}\frac{d^2L_{\Bar{u}}}{dh^2}\right]}\]

\section{Field theory calculations}

Here we give some details on the scaling forms
for various thermodynamic quantities arrived from
the spinless fermionic field theory written in the
theoretical models section in the main text. 
They can be found by known methods, e.g.
see Chapter 16 of Ref.~\cite{Sachdev_2011}.
One can do a scaling or power counting analysis
to show that the interaction term is RG-irrelevant 
based on dimensional grounds. This scaling argument is also
described in Sec. 16.2 of Ref.~\cite{Sachdev_2011}. 
Thus, we can arrive at the dominant scaling behavior for the thermodynamic
quantities by using the free field theory and ignoring the
interaction term.
The first result is for the overall free energy which takes
the following scaling form:
\begin{equation}
\mathcal{F}(T,\mu) = (k_B T) \left(\frac{k_B T}{J} \right)^{\frac{1}{2}} 
\; \Phi\left(\frac{\mu}{k_B T}\right) = J^{-1/2} \beta^{-3/2} \Phi(\beta \mu)
\end{equation}
where $\Phi(x)$ is an universal scaling function,
and $\mu \equiv 2 J - g \mu_B H$. $\beta = (k_B T)^{-1}$.
Using $C(\mu,T) = - \frac{1}{k_B T^2} \frac{\partial^2 (\beta F)}{\partial \beta^2}
$,
we get the following scaling form for the specific heat:
\begin{equation}
\frac{C}{k_B} = - \frac{3}{4} \left(\frac{k_B T}{J} \right)^{1/2} \Phi\left(\frac{\mu}{k_B T}\right) 
+ \left(\frac{\mu^2}{J (k_B T)} \right)^{1/2} \Phi'\left(\frac{\mu}{k_B T}\right)
- \left(\frac{\mu^4}{J (k_B T)^3} \right)^{1/2} \Phi''\left(\frac{\mu}{k_B T}\right)
\end{equation}
which can be re-expressed simply as
\begin{equation}
\frac{C}{k_B} = \sqrt{\frac{k_B T}{J}} \Theta \left(\frac{\mu}{k_B T} \right)
\end{equation}
where $\Theta(x)$ is an universal scaling function.
For the magnetization, one can calculate
deviation from the saturation magnetization corresponding to the fully polarized
state $|\ldots \downarrow \downarrow \downarrow \ldots\rangle$. This deviation becomes
the spinless fermionic occupation
given by
$\langle N \rangle = \langle \sum_i n_i \rangle = - \frac{\partial \mathcal{F}}{\partial \mu}$,
we get the following scaling form for the magnetization
\begin{equation}
M(H) - M_{\text{sat}} = g \mu_B \left(\frac{k_B T}{J} \right)^{\frac{1}{2}} 
\; \Phi'\left(\frac{\mu}{k_B T}\right)
\end{equation}
This overall scaling form can be further specified
in various regimes of the phase diagram. In the fully
polarized regime ($H > H_s$, $\mu < 0$, $k_B T \ll |\mu| \ll J$),
 we get
\begin{equation}
M(H) - M_{\text{sat}} = g \mu_B \left(\frac{k_B T}{4 \pi J} \right)^{\frac{1}{2}}
e^{\frac{\mu}{k_B T}}
\end{equation}
In the TLL regime ($H < H_s$, $\mu > 0$, $k_B T \ll |\mu| \ll J$),
which at quadratic level is a liquid of spinons at
finite density, we get 
\begin{equation}
M(H) - M_{\text{sat}} = g \mu_B \frac{1}{4 \pi (0.5)!}  
\left[ 
\left(\frac{\mu}{J}\right)^{\frac{1}{2}} \left(
1 - \frac{1}{24} \left( \frac{k_B T}{\mu} \right)^2
+ \ldots \right)
\right]
\end{equation}
In the QC regime ($H \sim H_s$, $|\mu| \ll k_B \ll J$) inside the "cone",
\begin{equation}
M(H) - M_{\text{sat}} = g \mu_B \zeta(0.5)(1-\sqrt{2}) \left( \frac{k_B T}{4\pi J} \right)^{\frac{1}{2}}
\end{equation}
which is now independent of $\mu$ as far as the
dominant behaviour goes when contrasted with the previous limits.

\FloatBarrier
